\documentclass[11pt,twoside]{article}
\usepackage{graphicx}
\usepackage{amsmath}
\usepackage{amssymb}
\usepackage{cite}
\usepackage{mathrsfs}
\usepackage{wrapfig}

\begin{document}

\newcommand{\pst}{\hspace*{1.5em}}

\newcommand{\rigmark}{\em Journal of Russian Laser Research}
\newcommand{\lemark}{\em Volume 31, Number 2, 2010}

\newcommand{\be}{\begin{equation}}
\newcommand{\ee}{\end{equation}}
\newcommand{\bm}{\boldmath}
\newcommand{\ds}{\displaystyle}
\newcommand{\bea}{\begin{eqnarray}}
\newcommand{\eea}{\end{eqnarray}}
\newcommand{\ba}{\begin{array}}
\newcommand{\ea}{\end{array}}
\newcommand{\arcsinh}{\mathop{\rm arcsinh}\nolimits}
\newcommand{\arctanh}{\mathop{\rm arctanh}\nolimits}
\newcommand{\bc}{\begin{center}}
\newcommand{\ec}{\end{center}}

\thispagestyle{plain}

\label{sh}


\begin{center} {\Large \bf
\begin{tabular}{c}
MuSR METHOD AND TOMOGRAPHIC PROBABILITY\\[-1mm]
REPRESENTATION OF SPIN STATES\\[-1mm]
\end{tabular}
 } \end{center}
\bigskip
\begin{center} {\bf
Yu. M. Belousov$^{1}$, S. N. Filippov$^{1*}$, V. N.
Gorelkin$^{1}$, and V. I. Man'ko$^{2}$}
\end{center}
\medskip
\begin{center}
{\it
$^{1}$Moscow Institute of Physics and Technology (State University)\\
Institutskii per. 9, Dolgoprudnyi, Moscow Region 141700, Russia\\
\smallskip
$^2$P.~N.~Lebedev Physical Institute, Russian Academy of Sciences\\
Leninskii Prospect 53, Moscow 119991, Russia}

\smallskip

*Corresponding author, e-mail:~~~ sergey.filippov@phystech.edu\\
e-mail:~~~ theorphys@phystech.edu,~~~ gorelkin2004@mail.ru,~~~
manko@sci.lebedev.ru
\end{center}

\begin{abstract}\noindent
Muon spin rotation/relaxation/resonance (MuSR) technique for
studying matter structures is considered by means of a recently
introduced probability representation of quantum spin states. A
relation between experimental MuSR histograms and muon spin
tomograms is established. Time evolution of muonium, anomalous
muonium, and a muonium-like system is studied in the tomographic
representation. Entanglement phenomenon of a bipartite
muon-electron system is investigated via tomographic analogues of
Bell number and positive partial transpose (PPT) criterion.
Reconstruction of the muon-electron spin state as well as the
total spin tomography of composed system is discussed.
\end{abstract}


\section{\label{section-introduction} Introduction}
\pst The description of spin states is usually based on the
concept of density matrix. The density matrix formalism was
introduced originally by Landau~\cite{landau} and von
Neumann~\cite{von-neumann} for continuous variables, and then
successfully extended to the case of discrete variables and spin
states (see, e.g.~\cite{fano}). In contrast to the state of
classical particle, the density matrix cannot be measured
immediately in the experiment, as the outcomes of quantum
observables have a probabilistic nature. The attempt to bring
classical and quantum descriptions closer together was made by
Wigner~\cite{wigner}, who introduced the quasi-probability
function $W(q,p)$ in phase space such that marginal distributions
$\int W(q,p) dp$ and $\int W(q,p) dq$ are true probability
distribution of position and momentum, respectively. However,
Wigner function can take negative values. Moreover, there is no
direct way to measure $W$-function. So far many quasi-probability
functions were introduced, for instance, Hisimi
$Q$-function~\cite{husimi} and Sudarshan-Glauber
$P$-function~\cite{sudarshan,glauber} (the review of different
quasi-probability distributions is presented in~\cite{vourdas}).
Experimental needs to reconstruct density matrix or Wigner
function of photon states by means of experimentally measured
probability distributions resulted in developing tomographic
methods (both theoretical~\cite{berber,vogel,mancini95,mancini}
and experimental~\cite{raymer,mlynek}) based on Radon
transform~\cite{radon,gelfand}. Though tomograms were originally
considered as auxiliary tools, recently it was proposed to treat
tomograms as primary notions of quantum
states~\cite{tombesi-manko,m-t-m-FondPhys} because a tomogram is a
fair probability distribution function, contains the same
information about the system as a density matrix, and the crucial
fact is that tomograms can be measured in laboratory. The
tomographic description of light states was extended to include
optical tomography~\cite{berber,vogel}, symplectic
tomography~\cite{mancini95}, photon-number
tomography~\cite{mancini,banaszek,wvogel}, and their recent
generalizations~\cite{asorey,asorey-arxiv}. The
tomographic-probability representation was then extended to spin
states~\cite{dodonovPLA,oman'ko-jetp} and developed
in~\cite{serg-spin,serg-chebyshev}. In the
paper~\cite{man'ko-sudarshan}, the general form of unitary spin
tomography was considered. The spin tomography with a finite
number of rotations was discussed in~\cite{serg-inverse-spin}. We
refer the reader to the review~\cite{ibort} for detailed history
of the tomographic-probability representation.

The aim of this paper is to apply spin tomograms to muon spin
rotation/relaxation/resonance experiments usually abbreviated to
MuSR or acronym $\mu$SR. At present, MuSR method is primarily used
to get insight about the microscopic behavior of materials and
nanostructures (see the
reviews~\cite{smilga-belousov,UFN79,UFN90}), although the origin
of this method is due to experimental verification of the proposal
that the weak interaction might violate parity symmetry, and
measuring the anisotropy of muon decay predicted by $V$-$A$ theory
of weak interactions. Muons ($\mu$) are utilized in $\mu$SR
technique as probes. Muon is a short-lived subatomic particle and
has the following properties: charge $\pm 1$ elementary charge
($\mu^{+}$ and $\mu^{-}$, respectively), mass $m_{\mu} =
206.77~m_e$, spin $s=1/2$, magnetic moment $\mu_{\mu} =
3.18334~\mu_p$, mean lifetime $\tau = 2.197 \cdot 10^{-6}~s$.
Positive (negative) muon decays into positron (electron) and a
neutrino-antineutrino pair as follows:
\begin{equation}
\mu^{+} \longrightarrow e^{+} + \nu_e + \tilde{\nu}_{\mu},
\qquad\qquad \mu^{-} \longrightarrow e^{-} + \tilde{\nu}_e +
{\nu}_{\mu}.
\end{equation}
Thanks to violation of parity symmetry, during the decay of
$\mu^{+}$, the positron $e^{+}$ is preferentially emitted along
the direction of muon spin. The decay positron can be emitted
along an arbitrary direction but the probability of such an event
depends on the direction ${\bf n} =
(\cos\varphi\sin\theta,\sin\varphi\sin\theta,\cos\theta)$
(anisotropy). Nevertheless, it is possible to measure the angular
distribution $\Gamma({\bf n})$ of the decay positrons from a bunch
of muons (ensemble) deposited at the same conditions. If detectors
are capable to detect positrons of an arbitrary energy (from 0 to
$\infty$) and muon spin is aligned along the $z$-axis, i.e. its
vector state $|j^{(\mu)}=1/2,m^{(\mu)}=1/2\rangle$ is an
eigenstate of angular momentum operators $\hat{J}_z$ and $\hat{\bf
J}^2$, then the angular distribution $\Gamma({\bf n})$ reads (see,
e.g.~\cite{smilga-belousov})
\begin{equation}
\label{gamma} \Gamma({\bf n})\frac{d{\bf n}}{4\pi} =
\left(1+\frac{1}{3}\cos\theta\right) \frac{\sin\theta d\theta
d\varphi}{4\pi}.
\end{equation}
\noindent The corresponding directional diagram is depicted in
Fig. \ref{figure1}a. The asymmetry of angular distribution
(\ref{gamma}) allows to reconstruct the spin state of muon.
Moreover, observing the time evolution $\Gamma({\bf n},t)$
(measuring $\mu$SR histograms) allows to trace the time evolution
of the muon spin polarization which depends sensitively on the
magnetic environment. In this paper, we relate the $\mu$SR
histogram $\Gamma({\bf n}, t)$ and a spin tomogram $w(m,{\bf
n},t)$.

The other aim of this paper is to consider the spin system of
muonium ${\rm Mu}=\mu^{+}e^{-}$ from the tomographic point of
view. Muonium is formed in certain materials (for instance, in
semiconductors) when a positive muon $\mu^{+}$ picks up an
electron. If this is the case, muonium turns out to be a light
isotope of hydrogen and is of interest not only in solid state
physics but also in chemistry~\cite{rhodes}. Thanks to a hyperfine
coupling, the information contained in the electron spin can be
passes on to the muon spin which is then analyzed via different
$\mu$SR techniques. Employing the tomographic-probability
representation, we are interested not only in reconstructing
density matrix of muon-electron spin state but also in formation,
controlling, and detecting entanglement of this system. Such an
approach would be useful to realize in practice a challenging
idea: check Bell inequalities with the help of true spins states
not polarizations of photons.

The paper is organized as follows.

In Sec.~\ref{section-muon-spin-tomography}, we review briefly the
tomography of a single spin and derive the relation between the
$\mu$SR histogram and the qubit tomogram. In
Sec.~\ref{section-two-spin-tomography}, we focus our attention on
muonium and muonium-like systems and address the problem of
two-spin tomography. In Sec.~\ref{section-entanglement-tomogram},
using the tomographic-probability representation, the entanglement
phenomenon, Bell-like inequality, and positive partial transpose
are discussed. In Sec.~\ref{section-evolution-tomogram}, a time
evolution of tomograms is considered. In
Sec.~\ref{section-Mu-hamiltonian}, we address muonium and
muonium-like systems in order to introduce their spin Hamiltonian
operator. In Sec.~\ref{section-tomography-Mu}, the time evolution
and entanglement of Mu and Mu-like systems are investigated in the
tomographic representation. Also, a method for reconstructing
muon-electron spin state is discussed. In
Sec.~\ref{section-conclusions}, conclusions are presented.

\section{\label{section-muon-spin-tomography} Muon Spin Tomography}
\pst Any spin-$j$ state given by the density operator $\hat{\rho}$
can be alternatively described by the following unitary spin
tomogram $w^{(j)}(m,\hat{u})$~\cite{man'ko-sudarshan}:
\begin{equation}
\label{unitary-spin-tomogram} w^{(j)}(m,\hat{u}) = \langle jm |
\hat{u} \hat{\rho} \hat{u}^{\dag} | jm \rangle,
\end{equation}
\noindent where $|jm\rangle$ is an eigenvector of angular momentum
operators $\hat{J}_z$ and $\hat{\bf J}^2$, $m$ is the spin
projection on $z$-direction ($m=-j,-j+1,\ldots,j$), $\hat{u}$ is a
general unitary transformation which is given by
$(2j+1)\times(2j+1)$ matrix $u \in SU(N)$, $N=2j+1$ in the basis
of states $|jm\rangle$. The inverse mapping $w^{(j)}(m,\hat{u})$
is also developed in~\cite{man'ko-sudarshan} and implies
integration over all unitary matrices $u\in SU(N)$ via a
corresponding measure. We emphasize the unique properties of the
unitary spin tomogram (\ref{unitary-spin-tomogram}):
non-negativity $w^{(j)}(m,\hat{u}) \ge 0$ and normalization
$\sum_{m=-j}^{j} w^{(j)}(m,\hat{u}) = 1$. The physical meaning of
the tomogram (\ref{unitary-spin-tomogram}) is the probability to
obtain the spin projection $m$ after a unitary rotation $\hat{u}
\hat{\rho} \hat{u}^{\dag}$ of the spin-$j$ state is fulfilled.

For our purposes, it is important to consider the case $u\in
SU(2)$, when the unitary spin tomogram transforms into a spin
tomogram $w(m,{\bf n})$ introduced
in~\cite{dodonovPLA,oman'ko-jetp}. Indeed, if this is the case, it
is possible to parameterize $\hat{u} = \hat{R}({\bf n})$ by unit
vector ${\bf n} =
(\cos\varphi\sin\theta,\sin\varphi\sin\theta,\cos\theta)$ as
follows:
\begin{equation}
\label{rotation-SU2} \hat{R}({\bf n}) = e^{- i ({\bf n}_{\bot}
\cdot \ \hat{\bf J}) \theta }, \qquad {\bf n}_{\bot} = (-\sin\phi,
\cos\phi, 0).
\end{equation}
\noindent For example, in case of qubits ($j=1/2$) the matrix of
operator (\ref{rotation-SU2}) in the basis of states $|jm\rangle$
is
\begin{equation}
R({\bf n}) = \left(%
\begin{array}{cc}
  \cos(\theta/2) & -\sin(\theta/2)e^{-i\phi} \\
  \sin(\theta/2)e^{i\phi} & \cos(\theta/2) \\
\end{array}%
\right).
\end{equation}
Thus, we obtain the spin tomogram $w(m,{\bf n}) = \langle jm |
\hat{R}({\bf n}) \hat{\rho} \hat{R}^{\dag}({\bf n}) | jm \rangle$
which has a sense of probability to obtain $m$ as a result of
measuring spin projection on direction ${\bf n}$. The inverse
mapping of spin tomogram $w^{(j)}(m, {\bf n})$ onto the density
operator $\hat{\rho}$ reads
\begin{equation}
\label{rho-reconstr-tomographic} \hat{\rho} =
\sum\limits_{m=-j}^{j} \frac{1}{4\pi} \int\limits_{0}^{2\pi}
d\varphi \int\limits_{0}^{\pi} \sin \theta d\theta ~
w^{(j)}(m,{\bf n}(\theta,\varphi)) \hat{D}^{(j)}(m,{\bf
n}(\theta,\varphi)).
\end{equation}
\noindent where $\hat{D}^{(j)}(m,{\bf n}(\theta,\varphi))$ is a
quantizer operator. According to~\cite{dodonovPLA,oman'ko-jetp},
the quantizer operator is expressed through Wigner $D$-function
$D_{m'm}^{(j)}(\alpha,\beta,\gamma)$ as follows:
\begin{eqnarray}
\label{quantizer-SU2-general} \hat{D}^{(j)}(m,{\bf
n}(\theta,\varphi)) &=& (-1)^{m_2'}
\sum_{j_3=0}^{2j}\sum_{m_3=-j_3}^{j_3} (2j_3+1)^2
\sum_{m_1,m_1',m_2' = -j}^{j}
(-1)^{m_1}D_{0m_3}^{(j_3)}(\theta,\varphi,\gamma) \nonumber\\
&& \times \left(%
\begin{array}{ccc}
  j & j & j_3 \\
  m_1 & -m_1 & 0 \\
\end{array}%
\right) \left(%
\begin{array}{ccc}
  j & j & j_3 \\
  m_1' & -m_2' & m_3 \\
\end{array}%
\right) |jm_1'\rangle\langle jm_2'|.
\end{eqnarray}
\noindent We cannot help mentioning that other different forms of
the quantizer operator are also known~\cite{serg-spin,castanos}.

\subsection{\label{subsection-tomography-of-muon-spin-state}Tomography of muon spin state}
\pst  As the tomographic technique is to be applied to muon spin,
let us consider the special case $j=1/2$. To avoid confusion, we
will denote muon spin $j^{(\mu)}=1/2$ rather than $j$, muon spin
projection $m^{(\mu)} = \pm 1/2$, and angular momentum operators
of muon spin are designated by $\hat{\bf J}^{(\mu)} \equiv
(\hat{J}_x^{(\mu)},\hat{J}_y^{(\mu)},\hat{J}_z^{(\mu)}) =
\frac{1}{2}\hat{\boldsymbol\sigma}$, where
$\hat{\boldsymbol\sigma} \equiv
(\hat{\sigma}_x,\hat{\sigma}_y,\hat{\sigma}_z)$ is the set of
Pauli operators.

Using definition (\ref{unitary-spin-tomogram}) and formula
(\ref{rotation-SU2}), it is not hard to see that the spin tomogram
of a single muon reads
\begin{equation}
w^{(\mu)}(m^{(\mu)},{\bf n}) = \frac{1}{2} + m^{(\mu)}  {\rm Tr}
\big[ \hat{\rho}^{(\mu)} ~ ({\bf n} \cdot \hat{\boldsymbol\sigma})
\big],
\end{equation}
\noindent whereas the quantizer operator can be written in the
following form~\cite{serg-spin}:
\begin{equation}
\hat{D}^{(\mu)} (m^{(\mu)},{\bf n}) = \frac{1}{2} \hat{I} + 3
m^{(\mu)} ({\bf n} \cdot \hat{\boldsymbol\sigma}),
\end{equation}
\noindent with $\hat{I}$ being the identity operator.

\begin{figure}
\begin{center}
\includegraphics{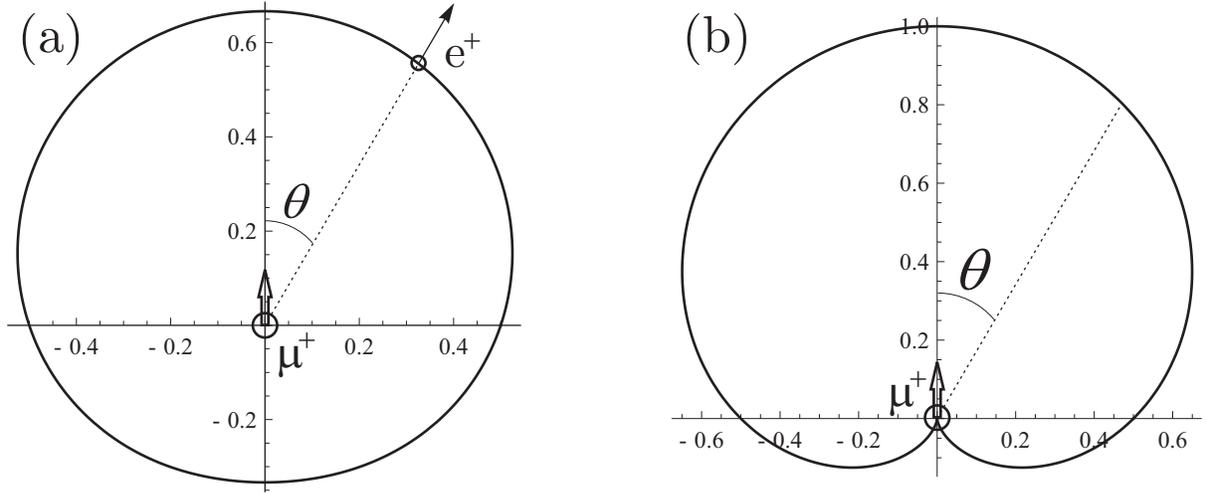}
\caption{\label{figure1} \small The relation between the angular
distribution $\Gamma({\bf n}(\theta,\varphi))$ of positrons
$e^{+}$ from muon $\mu^{+}$ decay (a) and the muon spin tomogram
$w^{(\mu)}(m^{(\mu)}=+1/2,{\bf n}(\theta,\varphi))$ of the state
$|\Uparrow\rangle = |j^{(\mu)}=1/2,m^{(\mu)}=1/2\rangle$
(schematically depicted by an arrow). }
\end{center}
\end{figure}

In order to compare the muon tomogram $w^{(\mu)}(m^{(\mu)},{\bf
n})$ and the angular distribution of decay positrons $\Gamma({\bf
n})$, let us consider the tomogram of the state $|\Uparrow\rangle
\equiv |j^{(\mu)}=1/2, m^{(\mu)} = +1/2 \rangle$. Indeed, in this
case we obtain
\begin{eqnarray}
\label{muon-UP-tomogram}
\rho_{\Uparrow}^{(\mu)} = \left(%
\begin{array}{cc}
  1 & 0 \\
  0 & 0 \\
\end{array}%
\right), \qquad w_{\Uparrow}^{(\mu)}(+1/2,{\bf n}(\theta,\varphi))
= \frac{1}{2} (1 + \cos\theta), \qquad
w_{\Uparrow}^{(\mu)}(-1/2,{\bf
n}(\theta,\varphi)) = \frac{1}{2} (1 - \cos\theta).\nonumber\\
\end{eqnarray}
\noindent The tomogram $w_{\Uparrow}^{(\mu)}(+1/2,{\bf
n}(\theta,\varphi))$ is depicted in Fig. \ref{figure1}b. Comparing
(\ref{gamma}) and (\ref{muon-UP-tomogram}), we derive the
following relation between the angular distribution of decay
positrons (averaged over all positron energies) and the muon
tomogram:
\begin{equation}
\label{relation} w^{(\mu)}(m^{(\mu)}=+1/2,{\bf n}) =
\frac{3}{2}\Gamma({\bf n}) - 1, \qquad
w^{(\mu)}(m^{(\mu)}=-1/2,{\bf n}) = 2 - \frac{3}{2}\Gamma({\bf
n}).
\end{equation}
\noindent The relations (\ref{relation}) are valid for an
arbitrary moment of time $t$ and they give a direct way of
transforming $\mu$SR histograms into muon spin tomograms
$w^{(\mu)}(m^{(\mu)},{\bf n},t)$. Moreover, it can be easily
checked that the relations (\ref{relation}) are adequate not only
for pure states (muon polarization $|{\bf P}|=1$) but also for an
arbitrary mixed state ($|{\bf P}|<1$). As far as negative muons
$\mu^{-}$ are concerned, the relation between tomograms and
angular distributions of decay electrons is readily obtained by
exchanging right-hand sides of Eqs. (\ref{relation}).

In the real experiment, we should take into account the spectral
responsivity of positron detectors, and this might result in
changing coefficients of relation (\ref{relation}) but linearity
of the relation is to be conserved. In particular, if there were
available detectors sensitive only to maximum positron energy
$E_{\max} = 52.83$~MeV, it would be possible to measure muon
tomogram directly because in this case
$w^{(\mu)}(m^{(\mu)}=+1/2,{\bf n}) = \Gamma_{E_{\max}}({\bf n})$
and $w^{(\mu)}(m^{(\mu)}=-1/2,{\bf n}) = 1-\Gamma_{E_{\max}}({\bf
n})$.

It is known that the muon $2\times 2$ density matrix depends on
three real parameters only (vector of polarization ${\bf P}$) and
can be written in the form $\hat{\rho} = \frac{1}{2}(\hat{I} +
({\bf P}\cdot\hat{\boldsymbol\sigma}))$. The muon tomogram
$w^{(\mu)}(m^{(\mu)},{\bf n})$ is a function on sphere ${\bf n}\in
S^2$ (and a discrete spin projection). Thus, the muon tomogram
contains full and even redundant information about the system.
Nevertheless, reconstruction of the density matrix by integration
over sphere (see formula (\ref{rho-reconstr-tomographic}))
exploits as much experimental information as possible. However, we
admit that the practical realization of such an experimental setup
(when detectors cover $4\pi$ steradian) is a challenge. To
eliminate this difficulty one can resort to spin tomography with a
finite number of rotations~\cite{serg-inverse-spin}. In case of
muons, such a tomography exploits only 3 probabilities
$w^{(\mu)}(m^{(\mu)}=+1/2,{\bf n}_k)$, $k=1,2,3$. As it is shown
in~\cite{serg-inverse-spin}, the inverse mapping onto the density
operator exists whenever triple product $({\bf n}_1\cdot[{\bf n}_2
\times {\bf n}_3]) \ne 0$ and reads
\begin{equation}
\label{qubit-reconstruction} \hat{\rho} = \frac{1}{2} \hat{I} +
\sum_{k=1}^{3} \left[ 2 w^{(\mu)}(m^{(\mu)}=+1/2,{\bf n}_k) - 1
\right] (\hat{\boldsymbol\sigma} \cdot {\bf l}_k),
\end{equation}
\noindent where vectors ${\bf l}_k$ form a dual basis with respect
to directions $\{{\bf n}_{k}\}_{k=1}^{3}$:
\begin{equation}
{\bf l}_1 = \frac{[{\bf n}_2 \times {\bf n}_3]}{({\bf n}_1 \cdot
[{\bf n}_2 \times {\bf n}_3])}, \quad {\bf l}_2 = \frac{[{\bf n}_3
\times {\bf n}_1]}{({\bf n}_1 \cdot [{\bf n}_2 \times {\bf
n}_3])}, \quad {\bf l}_3 = \frac{[{\bf n}_1 \times {\bf
n}_2]}{({\bf n}_1 \cdot [{\bf n}_2 \times {\bf n}_3])}.
\end{equation}
It is worth noting that, from experimental point of view, the
advantage of formula (\ref{qubit-reconstruction}) is that it does
not restrict us to using orthogonal directions only. Though if the
errors are presented in measured probabilities
$w^{(\mu)}(m^{(\mu)}=+1/2,{\bf n}_k)$, $k=1,2,3$, the
reconstructed density matrix is shown to be the least erroneous
when directions $\{{\bf n}_k\}_{k=1}^{3}$ are mutually
orthogonal~\cite{serg-inverse-spin}. In up-to-date experiments the
background fraction of detector clicks is usually less that 1\% of
the initial number of counts. At longer timescale $t\gg\tau_{\mu}$
the background noise does not allow to measure the muon tomogram
because the number of muon decays becomes exponentially small.

\section{\label{section-two-spin-tomography} Muonium and Two-Spin Tomography}
\pst Dealing with muons, we should not forget their physical
implementation and behavior in matter as well as a concrete
experimental setup for muon decay investigation. As muons are to
be stopped for measuring angular distribution of decay particles,
it is important to study behavior of muons in solid state
materials. On the other hand, the study of $\mu$SR histograms
(muon spin tomograms) gives us a useful information about the
matter in question. Special attention is paid to the investigation
of nanostructures, and this is why we do not consider muons in
gases and restrict ourselves to solid state materials.

\subsection{Muons in Matter}
\pst If negative muon $\mu^{-}$ is incorporated in material, it
behaves as a heavy electron and is readily captured by atoms to
$1s$ ground state. During its way `down', muon usually loses the
major part of its polarization. In $1s$ state, a space part of the
muon wave function is nonzero at an atomic nucleus. This fact may
lead to a nuclear capture of the negative muon. As a result,
negative muons in matter exhibit relatively small degree of
polarization and much shorter lifetime (substantially less than
$\tau_{\mu}$). It is the reason why negative muons ($\mu^{-}$SR
technique) are not as often used while investigating local matter
properties as positive muons. However, if a negative muon is in
$1s$ ground state inside an atom with zero-spin nucleus, the whole
system ``atom+$\mu^{-}$'' can be regarded as an acceptor center
(see, e.g.~\cite{mamedov-gorelkin-negative-muons}). The
pseudonucleus ``nucleus+$\mu^{-}$'' has spin $1/2$ and keeps a
part of muons initial polarization. For example, negative muons in
Si can form $^{\mu}Al$ acceptor centers. The effective moment of
the electron shell in such centers can take values greater than
$1/2$, for example, the case $j^{(e)} = 3/2$ is discussed in the
papers~\cite{baturin-gorelkin,gor-mam-bat}. Such systems are also
of interest for us and we will consider them in
Secs.~\ref{subsection-Mu-like-Hamiltonian} and
\ref{section-tomography-Mu}.

In contrast, in a condensed matter, positive muons $\mu^{+}$ are
repelled by atomic nuclei and are retarded gradually down to Bohr
velocities, and then to thermal velocities within the total time
interval $t\sim 10^{-10}$~s. The retarded muon can either come to
rest at an interstitial site of high symmetry between the lattice
ions (diamagnetic state) or pick up an electron (formation of
muonium ${\rm Mu} = \mu^{+}e^{-}$). A spur model of muonium
formation is discussed in~\cite{gor-sol-kon-bat}. In addition,
muon can also become a part of a chemical bound, for instance,
with oxygen ${\rm O}^{-}\mu^{+}$ similarly to ${\rm OH}$ bound. We
will focus our attention to muonium. It is worth noting that
muonium has almost the same Bohr radius and ionization potential
as hydrogen. Also, utilizing ultra-low energy muons (with energies
from $0.5$ to $30$ keV) and selecting their velocities, it is
possible to investigate muoniums formed at the implantation depths
from less than a nanometer to several hundred
nanometers~\cite{morenzoni04,morenzoni06}. This fact is crucial
for studying surfaces and thin
films~\cite{morenzoni09,london-depth-2010}.

Muonium is of special interest for us because it can be considered
as a two-spin system. Indeed, averaging over space part of the
common wave function, muon spin ($j^{(\mu)}=1/2$) and electron
spin ($j^{(e)}=1/2$) are coupled by hyperfine interaction
$\hat{H}_{\rm hf} = \hbar\omega_0(\hat{\bf J}^{(\mu)}\odot\hat{\bf
J}^{(e)}) \equiv \hbar\omega_0
(\hat{J}_x^{(\mu)}\otimes\hat{J}_x^{(e)} +
\hat{J}_y^{(\mu)}\otimes\hat{J}_y^{(e)} +
\hat{J}_z^{(\mu)}\otimes\hat{J}_z^{(e)})$, where $\omega_0 =
4453$~MHz.

The aim of the subsequent subsection is to consider two-spin
tomography and apply it to the spin state of muonium and
muonium-like systems (we will refer to the muon spin $1/2$
together with the arbitrary effective electron spin $j^{(e)}$ as
Mu-like system).

\subsection{Two-Spin Tomography}
\pst Let us consider a system comprising two spins: muon spin
$j^{(\mu)}$ and an effective electron spin $j^{(e)}$. The state of
the system is described by a $N\times N$ density operator
$\hat{\rho}$, where $N=(2j^{(\mu)}+1)(2j^{(e)}+1)$. If we
introduce a basis of system states $\{|i\rangle\}_{i=1}^{N}$, the
density operator can be written in the form of density matrix
$\rho_{ij} = \langle i | \hat{\rho} | j \rangle$. There are two
common ways to choose basis states:

(i) eigenvectors $|j^{(\mu)}m^{(\mu)}\rangle | j^{(e)}m^{(e)}
\rangle$ of individual angular momentum operator
$\hat{J}_z^{(\mu)} \otimes \hat{J}_z^{(e)}$;

(ii) eigenvectors $|LM\rangle$ of the total angular momentum
operators $\hat{L}_z$ and $\hat{\bf L}$, where the total spin $L$
can take values $L =
|j^{(\mu)}-j^{(e)}|,\ldots,j^{(\mu)}+j^{(e)}$.

Evidently, these two bases are related by the matrix $U_{\rm CG}$
composed of the Clebsch-Gordon coefficients
$C_{j^{(\mu)}m^{(\mu)}j^{(e)}m^{(e)}}^{LM} \equiv \langle
j^{(\mu)}m^{(\mu)},j^{(e)}m^{(e)} | LM \rangle$. For muonium
($j^{(\mu)}=1/2$, $j^{(e)}=1/2$) the matrix $U_{\rm CG}$ reads
\begin{equation}
\left(%
\begin{array}{l}
  |1,1\rangle \\
  |1,0\rangle \\
  |1,-1\rangle \\
  |0,0\rangle \\
\end{array}%
\right) = U_{\rm CG} \left(%
\begin{array}{l}
  |\frac{1}{2},\frac{1}{2}\rangle|\frac{1}{2},\frac{1}{2}\rangle \\
  |\frac{1}{2},\frac{1}{2}\rangle|\frac{1}{2},-\frac{1}{2}\rangle \\
  |\frac{1}{2},-\frac{1}{2}\rangle|\frac{1}{2},\frac{1}{2}\rangle \\
  |\frac{1}{2},-\frac{1}{2}\rangle|\frac{1}{2},-\frac{1}{2}\rangle \\
\end{array}%
\right), \qquad U_{\rm CG} = \left(%
\begin{array}{cccc}
  1 & 0 & 0 & 0 \\
  0 & \frac{1}{\sqrt{2}} & \frac{1}{\sqrt{2}} & 0 \\
  0 & 0 & 0 & 1 \\
  0 & \frac{1}{\sqrt{2}} & -\frac{1}{\sqrt{2}} & 0 \\
\end{array}%
\right).
\end{equation}
Consequently, in the probability representation muonium can be
described by two tomograms, namely, an individual two-spin
tomogram $w(m^{(\mu)},{\bf n}^{(\mu)},m^{(e)}{\bf n}^{(e)})$ and a
total two-spin tomogram $w^{(L)}(M,{\bf N})$ discussed below.

\subsubsection{\label{subsubsection-individual-2-spin-tom} Individual Two-Spin Tomogram}
\pst Extending the notion of spin tomogram
(\ref{unitary-spin-tomogram}) to the case of bipartite spin
system, we immediately obtain the individual unitary two-spin
tomogram defined through
\begin{equation}
\label{individual-UNITARY-two-spin-tomogram}
w(m^{(\mu)},m^{(e)},\hat{U}) = \langle j^{(\mu)}m^{(\mu)},
j^{(e)}m^{(e)} | \hat{U} ~ \hat{\rho} ~ \hat{U}^{\dag} |
j^{(\mu)}m^{(\mu)}, j^{(e)}m^{(e)} \rangle,
\end{equation}
\noindent where $\hat{U}$ is a general $N\times N$ unitary
transformation, $N=(2j^{(\mu)}+1)(2j^{(e)}+1)$. If $\hat{U}$ is
factorized into $\hat{U} = \hat{R}({\bf n}^{(\mu)}) \otimes
\hat{R}({\bf n}^{(e)})$, where the rotation operator $\hat{R}$ is
given by formula (\ref{rotation-SU2}), we readily obtain the
so-called individual two-spin tomogram of the form
\begin{equation}
\label{individual-two-spin-tomogram} w(m^{(\mu)},{\bf
n}^{(\mu)},m^{(e)},{\bf n}^{(e)}) = \langle j^{(\mu)}m^{(\mu)},
j^{(e)}m^{(e)} | \hat{R}({\bf n}^{(\mu)}) \otimes \hat{R}({\bf
n}^{(e)}) ~ \hat{\rho} ~ \hat{R}^{\dag}({\bf n}^{(\mu)}) \otimes
\hat{R}^{\dag}({\bf n}^{(e)}) | j^{(\mu)}m^{(\mu)}, j^{(e)}m^{(e)}
\rangle,
\end{equation}
\noindent which has a clear physical meaning: the individual
two-spin tomogram is the joint probability to obtain the value
$m^{(\mu)}$ in measuring the muon spin projection on direction
${\bf n}^{(\mu)}$ and the value $m^{(e)}$ in measuring the
electron spin projection on direction ${\bf n}^{(e)}$. Here,
measurements are assumed to be simultaneous. Though, in this case,
such measurements are not prohibited by quantum mechanics, it is
not evident how to realize them in practice. The reconstruction of
the density matrix of the whole system is
\begin{equation}
\hat{\rho} = \sum\limits_{m^{(\mu)}=-j^{(\mu)}}^{j^{(\mu)}}
\int\limits_{S^2} \frac{d{\bf n^{(\mu)}}}{4\pi}
\sum\limits_{m^{(e)}=-j^{(e)}}^{j^{(e)}} \int\limits_{S^2}
\frac{d{\bf n}^{(e)}}{4\pi} ~ w(m^{(\mu)},{\bf
n}^{(\mu)},m^{(e)},{\bf n}^{(e)})
\hat{D}^{(j^{(\mu)})}(m^{(\mu)},{\bf n}^{(\mu)}) \otimes
\hat{D}^{(j^{(e)})}(m^{(e)},{\bf n}^{(e)}),
\end{equation}
\noindent where the local quantizer operator $\hat{D}^{(j)}(m,{\bf
n})$ is given by formula (\ref{quantizer-SU2-general}).

As mentioned above, the practical realization of measuring the
individual spin-tomogram is a challenge. In the fact, the
$\mu$SR-method (discussed in
Sec.~\ref{section-muon-spin-tomography}) allows measuring the muon
part of the tomogram only. Using the probability interpretation of
the tomogram (\ref{individual-two-spin-tomogram}), we can say that
only the following marginal distribution is measurable via $\mu$SR
technique:
\begin{eqnarray}
\label{reduced-tomogram} \widetilde{w}(m^{(\mu)},{\bf n}^{(\mu)})
&=& \sum_{m^{(e)}=-j^{(e)}}^{j^{(e)}} w(m^{(\mu)},{\bf
n}^{(\mu)},m^{(e)}{\bf n}^{(e)}) = {\rm Tr} \big[ \hat{\rho} ~
\hat{R}^{\dag}({\bf n}^{(\mu)})| j^{(\mu)}m^{(\mu)}\rangle \langle
j^{(\mu)}m^{(\mu)} | \hat{R}({\bf n}^{(\mu)}) \otimes
\hat{I}^{(e)} \big] \nonumber\\
&=& \langle j^{(\mu)}m^{(\mu)} | \hat{R}({\bf n}^{(\mu)}) ~ {\rm
Tr}_{e}[\hat{\rho}] ~ \hat{R}^{\dag}({\bf n}^{(\mu)})|
j^{(\mu)}m^{(\mu)}\rangle,
\end{eqnarray}
\noindent which is nothing else but the muon spin tomogram of the
reduced density operator ${\rm Tr}_{e}[\hat{\rho}]$ (trace is
taken over electron spin states). Here, we used the following
resolution of identity:
\begin{equation}
\hat{I}^{(e)} = \sum_{m^{(e)}=-j^{(e)}}^{j^{(e)}}
\hat{R}^{\dag}({\bf n}^{(e)})| j^{(e)}m^{(e)}\rangle \langle
j^{(e)}m^{(e)} | \hat{R}({\bf n}^{(e)}),
\end{equation}
\noindent which is valid for any direction ${\bf n}^{(e)}$. Since
the summation over spin projections $m^{(e)}$ leads to exclusion
of ${\bf n}^{(e)}$ in the reduced tomogram
(\ref{reduced-tomogram}) (this phenomenon is also known as
non-signalling), we draw a conclusion that it is impossible to
reconstruct the muonium density matrix by measuring the reduced
tomogram (\ref{reduced-tomogram}) only.

However, if we consider the individual unitary two-spin tomogram
(\ref{individual-UNITARY-two-spin-tomogram}), we will see that the
reduced tomogram
\begin{equation}
\label{reduced-UNITARY-tomogram} \widetilde{w}(m^{(\mu)},\hat{U})
= \sum_{m^{(e)}=-j^{(e)}}^{j^{(e)}} w(m^{(\mu)},m^{(e)},\hat{U}) =
\langle j^{(\mu)}m^{(\mu)} | {\rm Tr}_{e}[\hat{U} ~ \hat{\rho} ~
\hat{U}^{\dag}] | j^{(\mu)}m^{(\mu)}\rangle
\end{equation}
\noindent does contain the full information about the muonium spin
state. The physical meaning of the distribution function
(\ref{reduced-UNITARY-tomogram}) is a probability to obtain the
muon spin projection on $z$-axis being equal to  $m^{(\mu)}$ after
a unitary rotation $\hat{U}$ is fulfilled over the common density
operator $\hat{\rho}$. In Sec.~\ref{subsection-reconstruction}, we
will outline how to reconstruct the muonium (and Mu-like) density
matrix by using measurable muon tomogram
(\ref{reduced-UNITARY-tomogram}), where the role of unitary
transformation $\hat{U}$ is played by an evolution operator
$\hat{\mathscr{U}}(t)$.

\subsubsection{Total Two-Spin Tomogram}
\pst From the mathematical point of view, the total unitary
two-spin tomogram $w^{(L)}(M,\hat{U})$ is merely a set of diagonal
elements of the transformed density operator in the alternative
basis of states $|LM\rangle$:
\begin{equation}
\label{total-UNITARY-2-spin} w^{(L)}(M,\hat{U}) = \langle LM |
\hat{U} ~ \hat{\rho} ~ \hat{U}^{\dag} | LM \rangle.
\end{equation}
Nonetheless, the tomogram $w^{(L)}(M,\hat{U})$ has a physical
meaning different from that of the individual unitary two-spin
tomogram (\ref{individual-UNITARY-two-spin-tomogram}). Actually,
$w^{(L)}(M,\hat{U})$ is the probability to find the system in
$L$-spin configuration and to obtain the total spin projection
onto $z$-direction equal to $M$ after the density operator
$\hat{\rho}$ experiences a unitary transformation $\hat{U}$.
Obviously, the tomogram $w^{(L)}(M,\hat{U})$ is normalized, i.e.
$\sum_{L}\sum_{M=-L}^{L} w^{(L)}(M,\hat{U}) = 1$. The total
unitary two-spin tomogram (\ref{total-UNITARY-2-spin}) contains
the same information as the individual unitary two-spin tomogram
(\ref{individual-UNITARY-two-spin-tomogram}) because both of them
are nothing else but diagonal representations of the rotated
density operator $\hat{U}\hat{\rho}\hat{U}^{\dag}$, with the only
difference being in the basis used: either
$\{|j^{(\mu)}m^{(\mu)}\rangle|j^{(e)}m^{(e)}\rangle\}$ or
$\{|LM\rangle\}$.

Similarly to the case of individual two-spin tomography, where we
specified the unitary transformation in the factorized form of
individual rotations, i.e. $\hat{U}_{\rm ind} = \hat{R}({\bf
n}^{(\mu)}) \otimes \hat{R}({\bf n}^{(e)})$, it is tempting to
introduce separate rotations of subspaces $\{|L'M'\rangle\}$ and
$\{|L''M''\rangle\}$ with $L' \ne L''$. For this reason, we
introduce the following operator:
\begin{equation}
\hat{U}_{\rm tot} = \hat{R}^{(L')}({\bf n}_{L'}) \oplus
\hat{R}^{(L'')}({\bf n}_{L''}) \oplus \ldots \equiv
\bigoplus\limits_{L=|j^{(\mu)}-j^{(e)}|}^{j^{(\mu)}+j^{(e)}}
\hat{R}^{(L)}({\bf n}_L),
\end{equation}
\noindent where by sign $\oplus$ we designate the direct sum. This
means that the matrix representation $U_{\rm tot}$ of the operator
$\hat{U}_{\rm tot}$ in the basis of states $\{|LM\rangle\}$ is
block-diagonal, with the $L$-th block being the
$(2L+1)\times(2L+1)$ matrix $R^{(L)}({\bf n}_{L})$ defined by
formula (\ref{rotation-SU2}).

It is of vital importance to trace how the problem of addition of
angular moments is related with individual and total two-spin
tomograms. The result is
\begin{equation}
\bigoplus\limits_{L=|j^{(\mu)}-j^{(e)}|}^{j^{(\mu)}+j^{(e)}}
\hat{R}^{(L)}({\bf N}) \equiv \hat{R}^{(L')}({\bf N}) \oplus
\hat{R}^{(L'')}({\bf N}) \oplus \ldots =
\hat{R}^{(j^{(\mu)})}({\bf N}) \otimes \hat{R}^{(j^{(e)})}({\bf
N})
\end{equation}
\noindent or in the matrix form
\begin{equation}
\bigoplus\limits_{L=|j^{(\mu)}-j^{(e)}|}^{j^{(\mu)}+j^{(e)}}
{R}^{(L)}({\bf N}) \equiv {R}^{(L')}({\bf N}) \oplus
{R}^{(L'')}({\bf N}) \oplus \ldots = U_{\rm CG}^{\dag} ~
{R}^{(j^{(\mu)})}({\bf N}) \otimes {R}^{(j^{(e)})}({\bf N}) ~
U_{\rm CG},
\end{equation}
\noindent where $U_{\rm CG}$ is a matrix of Clebsch-Gordon
coefficients.

Arguing in the same manner as in case of individual two-spin
tomogram (\ref{individual-two-spin-tomogram}), we introduce the
following total two-spin probability distribution function of the
form
\begin{equation}
\label{PDF-2-spin} f^{(L)}(M,{\bf N}) = \langle LM |
\bigoplus\limits_{L=|j^{(\mu)}-j^{(e)}|}^{j^{(\mu)}+j^{(e)}}
\hat{R}^{(L)}({\bf N}) ~ \hat{\rho} ~
\bigoplus\limits_{L=|j^{(\mu)}-j^{(e)}|}^{j^{(\mu)}+j^{(e)}}
\hat{R}^{(L)\dag}({\bf N}) | LM \rangle.
\end{equation}
\noindent It turns out that the knowledge of the probability
distribution function (\ref{PDF-2-spin}) is not enough, in
general, to reconstruct the system density operator of an
arbitrary mixed state~\cite{andreev-manko-jetp98}. Consequently,
we cannot regard the function $f^{(L)}(M,{\bf N})$ as a true
tomogram. However, if the information in (\ref{PDF-2-spin}) is
supplemented by probabilities with which pure states appear in the
mixed one, the mapping $\hat{\rho} \longrightarrow f^{(L)}(M,{\bf
N})$ is proved to be invertible~\cite{andreev-manko-jetp98}.

Despite the probability distribution (\ref{PDF-2-spin}) is not
informationally complete, it can still be treated as a tomogram of
identical spin-$j$ particles~\cite{dariano-paini}. Indeed, in case
of a bipartite system with indistinguishable constituents, the
density operator of the whole system $\rho$ is known to be
invariant with respect to a particle permutation. Hence, the
density matrix in the basis of states $\{|LM\rangle\}$ is
block-diagonal and can be reconstructed as follows:
\begin{equation}
\rho_{\rm ~indistinguishable} \equiv \bigoplus\limits_{L=0}^{2j}
\rho^{(L)} = \bigoplus\limits_{L=0}^{2j} ~ \sum\limits_{M=-L}^{L}
~ \int\limits_{S^2} \frac{d{\bf N}}{4\pi} ~ f^{(L)}(M,{\bf N})
\hat{D}^{(L)}(M,{\bf N}),
\end{equation}
\noindent where the quantizer $\hat{D}^{(L)}(M,{\bf N})$ is
determined by formula (\ref{quantizer-SU2-general}). As far as
other aspects of the addition of angular moments in the
tomographic-probability representation are beyond the MuSR-method,
they will be discussed elsewhere.

In order to conclude this consideration, we derive the relation
between the individual two-spin tomogram
(\ref{individual-two-spin-tomogram}) and the total unitary
two-spin tomogram (\ref{total-UNITARY-2-spin})
\begin{eqnarray}
w^{(L)}(M,\hat{U}) =
\sum\limits_{m^{(\mu)}=-j^{(\mu)}}^{j^{(\mu)}} \int\limits_{S^2}
\frac{d{\bf n^{(\mu)}}}{4\pi}
\sum\limits_{m^{(e)}=-j^{(e)}}^{j^{(e)}} \int\limits_{S^2}
\frac{d{\bf n^{(e)}}}{4\pi} ~ w(m^{(\mu)},{\bf
n}^{(\mu)},m^{(e)},{\bf n}^{(e)}) \nonumber\\
\times \langle LM | \hat{U} ~ \hat{D}^{(j^{(\mu)})}(m^{(\mu)},{\bf
n}^{(\mu)}) \otimes \hat{D}^{(j^{(e)})}(m^{(e)},{\bf n}^{(e)}) ~
\hat{U}^{\dag} | LM \rangle.
\end{eqnarray}

\section{\label{section-entanglement-tomogram} Entanglement and Separability in Probability Representation}
\pst A state of two-spin system (muon and electron spins) is
called separable if the density operator $\hat{\rho}$ of the whole
system can be resolved into the following convex sum
\begin{equation}
\label{separability} \hat{\rho}_{\rm separable} = \sum_{i} p_i ~
\hat{\rho}_i^{(\mu)} \otimes \hat{\rho}_i^{(e)}, \qquad 0 \le p_i
\le 1, \quad \sum_i p_i =1,
\end{equation}
\noindent otherwise the state is called entangled. Since the
individual two-spin tomogram $w(m^{(\mu)},{\bf
n}^{(\mu)},m^{(e)},{\bf n}^{(e)})$ contains the same information
about the system involved, the definition of separability
(\ref{separability}) is readily reformulated in terms of
tomograms. In fact, the state is separable when
\begin{equation}
\label{tomographic-separability} w_{\rm separable}(m^{(\mu)},{\bf
n}^{(\mu)},m^{(e)},{\bf n}^{(e)}) = \sum_{i} p_i ~
w_i^{(j^{(\mu)})}(m^{(\mu)},{\bf n}^{(\mu)})
w_i^{(j^{(e)})}(m^{(e)},{\bf n}^{(e)}).
\end{equation}
\noindent If the resolution (\ref{tomographic-separability}) is
not fulfilled for any convex parameters $p_i$, then the state is
entangled.

\subsection{Entanglement Detection via Bell-like Inequality Violation}
\pst To detect the entanglement of a two-qubit system it is
convenient to check the violation of the following Bell-like
inequality (see, e.g.~\cite{chernega}):
\begin{equation}
\label{bell-like} |B| \le 2,
\end{equation}
\noindent where the Bell-like number $B$ reads
\begin{eqnarray}
B = {\rm Tr} \left[ \mathscr{I} \left(%
\begin{array}{cccc}
  w(+ ,{\bf n}_1^{(\mu)},+ ,{\bf n}_1^{(e)}) & w(+ ,{\bf n}_1^{(\mu)},+ ,{\bf n}_2^{(e)}) & w(+ ,{\bf n}_2^{(\mu)},+ ,{\bf n}_1^{(e)}) & w(+ ,{\bf n}_2^{(\mu)},+ ,{\bf n}_2^{(e)}) \\
  w(+ ,{\bf n}_1^{(\mu)},- ,{\bf n}_1^{(e)}) & w(+ ,{\bf n}_1^{(\mu)},- ,{\bf n}_2^{(e)}) & w(+ ,{\bf n}_2^{(\mu)},- ,{\bf n}_1^{(e)}) & w(+ ,{\bf n}_2^{(\mu)},- ,{\bf n}_2^{(e)}) \\
  w(- ,{\bf n}_1^{(\mu)},+ ,{\bf n}_1^{(e)}) & w(- ,{\bf n}_1^{(\mu)},+ ,{\bf n}_2^{(e)}) & w(- ,{\bf n}_2^{(\mu)},+ ,{\bf n}_1^{(e)}) & w(- ,{\bf n}_2^{(\mu)},+ ,{\bf n}_2^{(e)}) \\
  w(- ,{\bf n}_1^{(\mu)},- ,{\bf n}_1^{(e)}) & w(- ,{\bf n}_1^{(\mu)},- ,{\bf n}_2^{(e)}) & w(- ,{\bf n}_2^{(\mu)},- ,{\bf n}_1^{(e)}) & w(- ,{\bf n}_2^{(\mu)},- ,{\bf n}_2^{(e)}) \\
\end{array}%
\right) \right], \nonumber\\
\end{eqnarray}
\noindent where the sign $\pm$ denotes spin projection $\pm 1/2$,
respectively, and the matrix $\mathscr{I}$ is defined through
\begin{equation}
\label{I-matrix}
\mathscr{I} = \left(%
\begin{array}{cccc}
  1 & -1 & -1 & 1 \\
  1 & -1 & -1 & 1 \\
  1 & -1 & -1 & 1 \\
  -1 & 1 & 1 & -1 \\
\end{array}%
\right).
\end{equation}
If state is separable, then the inequality (\ref{bell-like}) is
necessarily valid. Thus, its violation is a direct evidence of the
system entanglement. Note, that we do not resort to the density
matrix formalism while dealing with such an approach.

\subsection{Positive Partial Transpose in the Tomographic-Probability Representation}
\pst Two-qubit and qubit-qutrit states are known to be separable
iff positive partial transpose $\rho^{\rm ppt}$ of the system
density matrix $\rho$ results in a new density matrix
(Peres-Horodecki criterion~\cite{peres,horodecki}). If $\rho^{\rm
ppt}$ is not a density matrix, this is the direct evidence of the
entanglement. The Peres-Horodecki criterion differs substantially
from the approach outlined in the previous subsection because it
provides the necessary and sufficient condition for a two-qubit
state or qubit-qutrit state to be entangled whereas the violation
of the Bell-like inequality (\ref{bell-like}) is only a sufficient
but not necessary indicator of entanglement.

Positive partial transpose (PPT) of bipartite state $\rho$ leaves
the second subsystem undisturbed and makes a transpose of the
first subsystem, which is equivalent to the complex conjugation of
the first subsystem (time reversion). Since the complex
conjugation of qubit density matrix $\rho = \frac{1}{2}(I+({\bf
P}\cdot\boldsymbol{\sigma}))$ boils down to the replacement $P_y
\rightarrow -P_y$, the corresponding qubit tomogram $w(m,{\bf n})$
transforms into $w(m,{\bf n}^{\rm ppt})$, where ${\bf n}^{\rm ppt}
\equiv (n_x,-n_y,n_z)$. Consequently, in case of two qubits
(muonium) and qubit-qutrit system (muonium-like system), the
action of positive partial transpose on the individual two-spin
tomogram (\ref{individual-two-spin-tomogram}) is
\begin{equation}
w^{\rm ppt}(m^{(\mu)},{\bf n}^{(\mu)},m^{(e)},{\bf n}^{(e)}) =
w(m^{(\mu)},{\bf n}^{(\mu){\rm ppt}},m^{(e)},{\bf n}^{(e)}).
\end{equation}
In the density-matrix formalism, the following step is to check a
positivity of the matrix $\rho^{\rm ppt}$. It is shown
in~\cite{serg-geometric} that eigenvalues
$\{\lambda_i\}_{i=1}^{4}$ of a $4\times 4$ Hermitian matrix
$\Lambda$ with unit trace (${\rm Tr}\Lambda = 1$) are non-negative
iff $M_{2} \equiv \sum\limits_{i < j} \lambda_i \lambda_j \ge 0$,
$M_{3} \equiv \sum\limits_{i < j < k} \lambda_i \lambda_j
\lambda_k \ge 0$, and $M_4 \equiv \sum\limits_{i < j < k < l}
\lambda_i \lambda_j \lambda_k \lambda_l \ge 0$. Using the trace
equality $\sum\limits_{i} \lambda_i = 1$, these quantities can be
expressed through traces ${\rm Tr}\Lambda^{n}$ as follows:
\begin{eqnarray}
&& \label{M-2} M_2 = \frac{1}{2} \left( 1 - {\rm Tr}\Lambda^2 \right) \ge 0, \\
&& \label{M-3} M_3 = \frac{1}{6} \left( 1 - 3{\rm Tr}\Lambda^2 + 2{\rm Tr}\Lambda^3 \right) \ge 0, \\
&& \label{M-4} M_4 = \frac{1}{24} \left( 1 - 6{\rm Tr}\Lambda^2 +
3 \big[{\rm Tr}\Lambda^2 \big]^2 + 8{\rm Tr}\Lambda^3 -6{\rm
Tr}\Lambda^4 \right) \ge 0.
\end{eqnarray}
Substituting $\rho^{\rm ppt}$ for $\Lambda$ in
(\ref{M-2})--(\ref{M-4}), it is possible to detect unambiguously
the entanglement of two-qubit system by observing a violation of
at least one inequality. The value of $M_2$ turns out to be
invariant with respect to PPT transformation. In view of this, we
introduce the following entanglement measure $E$ for two qubits:
\begin{equation}
\label{measure} E = |M_3| + |M_4| - M_3 - M_4.
\end{equation}
In the tomographic-probability approach, the values of $M_3$ and
$M_4$ read
\begin{eqnarray}
\label{M-3-tomographic} M_3 &=& \frac{1}{6} \sum_{m^{(\mu)}}\sum_{m^{(e)}} [ w^{\rm ppt} - 3w^{\rm ppt} \star w^{\rm ppt} + 2w^{\rm ppt} \star w^{\rm ppt} \star w^{\rm ppt}](m^{(\mu)},{\bf n}^{(\mu)},m^{(e)},{\bf n}^{(e)}), \qquad\\
\label{M-4-tomographic} M_4 &=& \frac{1}{24} \Bigg( 3 \Big(
\sum_{m^{(\mu)}}\sum_{m^{(e)}} [w^{\rm ppt} \star w^{\rm ppt}]
(m^{(\mu)},{\bf n}^{(\mu)},m^{(e)},{\bf n}^{(e)}) \Big)^2 +
\sum_{m^{(\mu)}}\sum_{m^{(e)}} [ w^{\rm ppt} - 6w^{\rm ppt} \star
w^{\rm ppt}  \nonumber\\
&& + 8w^{\rm ppt} \star w^{\rm ppt} \star w^{\rm ppt} - 6w^{\rm
ppt} \star w^{\rm ppt} \star w^{\rm ppt} \star w^{\rm
ppt}](m^{(\mu)},{\bf n}^{(\mu)},m^{(e)},{\bf n}^{(e)})  \Bigg),
\end{eqnarray}
\noindent where by $\star$ we denote a star product with the
kernel
\begin{eqnarray}
&& K(m_1'{\bf n}_1'm_2'{\bf n}_2',m_1''{\bf n}_1''m_2''{\bf
n}_2'', m_1 {\bf n}_1 m_2 {\bf n}_2) = \prod_{i=1}^2 \Big[
\frac{1}{4} + 9m_{i}'m_{i}'' \left( {\bf n}_{i}' \cdot {\bf
n}_{i}'' \right) + 3m_{i}m_{i}' \left( {\bf n}_{i} \cdot {\bf
n}_{i}' \right) \nonumber\\
&& + 3m_{i}m_{i}'' \left( {\bf n}_{i} \cdot {\bf n}_{i}'' \right)
+ i 18 m_{i}m_{i}'m_{i}'' \left( {\bf n}_{i} \cdot [{\bf n}_{i}'
\times {\bf n}_{i}''] \right) \Big].
\end{eqnarray}
Thus, the measure (\ref{measure}) is expressed through spin
tomograms only.

\section{\label{section-evolution-tomogram} Evolution of Tomograms}
\pst Evolution of pure states $|\psi\rangle$ is governed by
Schr\"{o}dinger equation $i\hbar \frac{\partial
|\psi\rangle}{\partial t} = \hat{H} |\psi\rangle$, where $\hat{H}$
is a Hamiltonian of the system. If we consider generally mixed
state with the density operator $\hat{\rho}$, the evolution obeys
the von Neumann equation $i\hbar \frac{\partial
\hat{\rho}}{\partial t} = [\hat{H},\hat{\rho}]$ which is similar
to Liuville equation in classical statistical mechanics. Taking
advantage of mapping $\hat{\rho}(t) \longrightarrow
w(m,\hat{u},t)$ for a single spin-$j$ particle (Eq.
(\ref{unitary-spin-tomogram})), the von Neumann equation with the
initial condition takes the form
\begin{equation}
\label{evolution-of-tomograms} i\hbar \frac{\partial
w(m,\hat{u},t)}{\partial t} = [ f_{H}\star w - w \star f_{H} ]
(m,\hat{u},t), \qquad w(m,\hat{u},t=0) = w_0(m,\hat{u}),
\end{equation}
\noindent where $f_{H}(m,\hat{u},t)$ is a tomographic symbol of
operator $\hat{H}$ obtained by replacing $\hat{\rho}\rightarrow
\hat{H}$ in formula (\ref{unitary-spin-tomogram}); a star $\star$
implies the so-called star product of tomographic symbols. In
other words,
\begin{equation}
[f_{H}\star w] (m,\hat{u},t) = \sum_{m',m''=-j}^{j} \int d\hat{u}'
\int d\hat{u}'' f_{H}(m',\hat{u}',t) f_{H}(m',\hat{u}',t)
K(m',\hat{u}',m'',\hat{u}'',m,\hat{u}),
\end{equation}
\noindent where $\int d\hat{u}$ is an integration over the
corresponding measure, $K(m',\hat{u}',m'',\hat{u}'',m,\hat{u})$ is
the star-product kernel of the form $K= Tr \big[
\hat{D}(m',\hat{u}') \hat{D}(m'',\hat{u}'') \hat{U}(m,\hat{u})
\big]$, where $\hat{D}$ and $\hat{U}$ are quantizer and
dequantizer operators, respectively. Explicit form of the spin
tomographic star-product kernel is found
in~\cite{serg-spin,serg-chebyshev,castanos}.

Formal solutions of Schr\"{o}dinger and von Neumann equations are
expressed through the unitary evolution operator
$\hat{\mathscr{U}}(t)$ as follows:
\begin{eqnarray}
\label{psi-rho-t} && |\psi(t)\rangle = \hat{\mathscr{U}}(t)
|\psi(0)\rangle, \qquad \hat{\rho}(t)
= \hat{\mathscr{U}}(t) \hat{\rho}(0) \hat{\mathscr{U}}^{\dag}(t),\\
&& \hat{\mathscr{U}}(t) = \hat{T}\exp \left[ - \frac{i}{\hbar}
\int_{0}^{t} \hat{H}(\tau) d\tau \right],
\end{eqnarray}
\noindent where $\hat{T}$ is a time-ordering operator.

Using the relation (\ref{psi-rho-t}) as well as the mapping
(\ref{unitary-spin-tomogram}), it is not hard to see that the time
evolution of a unitary spin tomogram reads
\begin{equation}
w(m,\hat{u},t) = w_0(m,\hat{u}\hat{\mathscr{U}}(t)).
\end{equation}
\noindent which is nothing else but the formal solution of the
tomographic evolution equation (\ref{evolution-of-tomograms}).

As far as muonium and muonium-like systems are concerned, the
evolution of the individual unitary two-spin tomogram
(\ref{individual-UNITARY-two-spin-tomogram}) reads
\begin{equation}
w(m^{(\mu)},m^{(e)},\hat{U},t) =
w_0(m^{(\mu)},m^{(e)},\hat{U}\hat{\mathscr{U}}(t)),
\end{equation}
\noindent whereas the evolution of the individual two-spin
tomogram (\ref{individual-two-spin-tomogram}) has a rather
complicated form
\begin{eqnarray}
\label{evolution-individual-tomogram} &&w(m^{(\mu)},{\bf
n}^{(\mu)},m^{(e)},{\bf n}^{(e)},t)
\nonumber\\
&&=\!\!\!\!\!\sum\limits_{\widetilde{m}^{(\mu)}=-j^{(\mu)}}^{j^{(\mu)}}
\int\limits_{S^2} \frac{d\widetilde{\bf n}^{(\mu)}}{4\pi}
\!\!\!\!\!\sum\limits_{\widetilde{m}^{(e)}=-j^{(e)}}^{j^{(e)}}
\int\limits_{S^2} \frac{d\widetilde{\bf n}^{(e)}}{4\pi} ~
w_0(\widetilde{m}^{(\mu)},\widetilde{\bf
n}^{(\mu)},\widetilde{m}^{(e)},\widetilde{\bf n}^{(e)}) ~ \langle
j^{(\mu)}m^{(\mu)}, j^{(e)}m^{(e)} | \hat{R}({\bf
n}^{(\mu)}) \otimes \hat{R}({\bf n}^{(e)})\nonumber\\
&& \times \hat{\mathscr{U}}(t)
\hat{D}^{(j^{(\mu)})}(\widetilde{m}^{(\mu)},\widetilde{\bf
n}^{(\mu)}) \otimes
\hat{D}^{(j^{(e)})}(\widetilde{m}^{(e)},\widetilde{\bf n}^{(e)})
\hat{\mathscr{U}}^{\dag}(t) ~ \hat{R}^{\dag}({\bf n}^{(\mu)})
\otimes \hat{R}^{\dag}({\bf n}^{(e)}) | j^{(\mu)}m^{(\mu)},
j^{(e)}m^{(e)} \rangle.
\end{eqnarray}


\section{\label{section-Mu-hamiltonian} Muonium and Mu-like Systems: Hamiltonian}
\pst Hamiltonian of a muonium-like system has the following most
general form:
\begin{equation}
\hat{H}_{\rm general}(q,s) = \hat{H}_0(q) + \hat{V}(q,s),
\end{equation}
\noindent where $q$ and $s$ are responsible for position and spin
variables, respectively; $\hat{H}_0(q)$ is a part depending on
space coordinates only; and $\hat{V}(q,s)$ is a part containing
spin variables. For muonium-like systems, Coulomb and spin-orbital
interactions are much greater than the hyperfine interaction ($H_0
\gg V$). For this reason, in our case the operator $\hat{V}(q,s)$
will only contain the hyperfine interaction and an interaction of
magnetic moments $\mu_{\mu}$ and $\mu_{e}$ with local magnetic
fields. On averaging over space coordinates, the operator
$\langle\psi(q)|\hat{V}(q,s)|\psi(q)\rangle$ acts on spin degrees
of freedom, i.e., it can be represented by $N\times N$
spin-Hamiltonian operator $\hat{H}$ with
$N=(2j^{(\mu)}+1)(2j^{(e)}+1)$.

\subsection{Muonium Hamiltonian}
\pst We will consider three main kinds of muonium Hamiltonian:
\begin{itemize}
    \item Free muonium atom in vacuum is governed by hyperfine
    interaction
    \begin{equation}
    \label{hyperfine-Hamiltonian}
    \hat{H}_{\rm hf} = \hbar\omega_0(\hat{\bf J}^{(\mu)}\odot\hat{\bf
J}^{(e)}) \equiv \hbar\omega_0
(\hat{J}_x^{(\mu)}\otimes\hat{J}_x^{(e)} +
\hat{J}_y^{(\mu)}\otimes\hat{J}_y^{(e)} +
\hat{J}_z^{(\mu)}\otimes\hat{J}_z^{(e)}),
    \end{equation}
    \noindent where the frequency of hyperfine interaction $\omega_0 =
    4453$~MHz. It is worth mentioning that such a Hamiltonian
    provides an adequate description of muonium not only in vacuum
    but also in quartz and ice. Frequency of hyperfine splitting in such materials is almost the same as in vacuum within the accuracy of the
    experiment. The correction caused by interaction of the electron quadrupole electrical
    moment with inhomogeneous crystal field can lead to
    anisotropic Hamiltonian (in quartz at temperatures $T \lesssim 100$~K).
    \item Isotropic Hamiltonian of a muonium-like system in solids in presence of magnetic
    field ${\bf B}$
    \begin{equation}
    \label{isotropic-Hamiltonian}
    \hat{H}_{\rm Mu} = A(\hat{\bf J}^{(\mu)}\odot\hat{\bf
J}^{(e)}) - g_{\mu}\mu_{\mu} ({\bf B}\cdot \hat{\bf J}^{(\mu)})
\otimes \hat{I}^{(e)} + g_e \mu_e \hat{I}^{(\mu)}\otimes ({\bf
B}\cdot\hat{\bf J}^{(e)}),
    \end{equation}
    \noindent where in general $A \ne \hbar\omega_0$, $g_{\mu}=2$, $\mu_e$ is the Bohr magneton, and $({\bf B}\cdot \hat{\bf J}) \equiv B_x\hat{J}_x + B_y\hat{J}_y +
    B_z\hat{J}_z$.
    \item Anisotropic Hamiltonian of anomalous muonium (Mu$^{\ast}$) which is
    present, for instance, in Si and many other semiconductors
    with a diamond structure of crystal lattice
    \begin{equation}
    \label{anisotropic-Hamiltonian}
\hat{H}_{\rm Mu^{\ast}} = \hat{H}_{\rm Mu} + \Delta A
(\boldsymbol{\mathfrak{N}}\cdot \hat{\bf J}^{(\mu)}) \otimes
(\boldsymbol{\mathfrak{N}}\cdot \hat{\bf J}^{(e)}),
    \end{equation}
\noindent where $\boldsymbol{\mathfrak{N}}$ is a unit vector
determining the axis of axial asymmetry.
\end{itemize}

If the Hamiltonian is time-independent (magnetic field is steady),
then the corresponding unitary evolution operators can be found
analytically. In the basis of states $| j^{(\mu)}m^{(\mu)},
j^{(e)}m^{(e)} \rangle$, the unitary evolution matrix of hyperfine
interaction reads
\begin{equation}
\label{unitary-evolution-hf}
\mathscr{U}_{\rm hf} (t) = \frac{1}{2}e^{-i\omega_0 t/4}\left(%
\begin{array}{cccc}
  2 & 0 & 0 & 0 \\
  0 & 1+e^{i \omega_0 t} & 1-e^{i \omega_0 t} & 0 \\
  0 & 1-e^{i \omega_0 t} & 1+e^{i \omega_0 t} & 0 \\
  0 & 0 & 0 & 2 \\
\end{array}%
\right)
\end{equation}
In order to find the evolution matrix of Mu in presence of
magnetic field, the intensity ${\bf B}$ is chosen to be aligned
along $z$-direction, i.e. ${\bf B} \parallel z$ (longitudinal
field). In this case, we obtain
\begin{equation}
\label{unitary-evolution-hf-field}
\mathscr{U}_{{\rm Mu}~z} (t) = e^{iat} \left(%
\begin{array}{cccc}
  e^{-i(2a-b_{-})t} & 0 & 0 & 0 \\
  0 & \cos ct + i (b_{+}/c) \sin ct & -i (2a/c) \sin ct & 0 \\
  0 & -i (2a/c) \sin ct & \cos ct - i (b_{+}/c) \sin ct & 0 \\
  0 & 0 & 0 & e^{-i(2a+b_{-})t} \\
\end{array}%
\right),
\end{equation}
\noindent where $a=A/(4\hbar)$, $b_{\pm} = B(g_{\mu}\mu_{\mu} \pm
g_{e}\mu_{e})/(2\hbar)$, and $c=\sqrt{A^2+B^2(g_{\mu}\mu_{\mu} +
g_{e}\mu_{e})^2}/(2\hbar)$.

If the magnetic field is aligned along $x$-axis (transversal
field), the unitary evolution matrix $\mathscr{U}_{{\rm Mu}~x}
(t)$ is symmetrical, and its matrix elements read
\begin{eqnarray}
&& \label{unitary-evolution-hf-field-x-11}[\mathscr{U}_{{\rm
Mu}~x} (t)]_{11} = [\mathscr{U}_{{\rm Mu}~x} (t)]_{44} =
\frac{1}{2} \left[ e^{-iat} \cos b_{-}t + e^{iat}
\left( \cos ct - i (2a/c) \sin ct \right) \right], \qquad\qquad \\
&& [\mathscr{U}_{{\rm Mu}~x} (t)]_{22} = [\mathscr{U}_{{\rm Mu}~x}
(t)]_{33} = \frac{1}{2} \left[ e^{-iat} \cos b_{-}t + e^{iat}
\left( \cos ct + i (2a/c) \sin ct \right) \right], \\
&& [\mathscr{U}_{{\rm Mu}~x} (t)]_{12} = [\mathscr{U}_{{\rm Mu}~x}
(t)]_{34} = -\frac{i}{2} \left[ e^{-iat} \sin b_{-}t + (b_{+}/c)
e^{iat}
\sin ct \right], \\
&& [\mathscr{U}_{{\rm Mu}~x} (t)]_{13} = [\mathscr{U}_{{\rm Mu}~x}
(t)]_{24} = -\frac{i}{2} \left[ e^{-iat} \sin b_{-}t - (b_{+}/c)
e^{iat}
\sin ct \right], \\
&& [\mathscr{U}_{{\rm Mu}~x} (t)]_{14} = \frac{1}{2} \left[
e^{-iat} \cos b_{-}t - e^{iat}
\left( \cos ct - i (2a/c) \sin ct \right) \right], \\
&& \label{unitary-evolution-hf-field-x-23}[\mathscr{U}_{{\rm
Mu}~x} (t)]_{23} = \frac{1}{2} \left[ e^{-iat} \cos b_{-}t -
e^{iat} \left( \cos ct + i (2a/c) \sin ct \right) \right].
\end{eqnarray}
Finally, if direction of the magnetic field coincides with
$y$-axis, the unitary evolution matrix $\mathscr{U}_{{\rm Mu}~y}
(t)$ is expressed through matrix $\mathscr{U}_{{\rm Mu}~x} (t)$ as
follows:
\begin{equation}
\label{unitary-evolution-hf-field-y}
\mathscr{U}_{{\rm Mu}~y} (t) = \left(%
\begin{array}{cccc}
  [\mathscr{U}_{{\rm Mu}~x} (t)]_{11} & -i[\mathscr{U}_{{\rm Mu}~x} (t)]_{12} & -i[\mathscr{U}_{{\rm Mu}~x} (t)]_{13} & -[\mathscr{U}_{{\rm Mu}~x} (t)]_{14} \\
  i[\mathscr{U}_{{\rm Mu}~x} (t)]_{21} & [\mathscr{U}_{{\rm Mu}~x} (t)]_{22} & [\mathscr{U}_{{\rm Mu}~x} (t)]_{23} & -i[\mathscr{U}_{{\rm Mu}~x} (t)]_{24} \\
  i[\mathscr{U}_{{\rm Mu}~x} (t)]_{31} & [\mathscr{U}_{{\rm Mu}~x} (t)]_{32} & [\mathscr{U}_{{\rm Mu}~x} (t)]_{33} & -i[\mathscr{U}_{{\rm Mu}~x} (t)]_{34} \\
  -[\mathscr{U}_{{\rm Mu}~x} (t)]_{41} & i[\mathscr{U}_{{\rm Mu}~x} (t)]_{42} & i[\mathscr{U}_{{\rm Mu}~x} (t)]_{43} & [\mathscr{U}_{{\rm Mu}~x} (t)]_{44} \\
\end{array}%
\right).
\end{equation}
As concerns the anisotropic hyperfine interaction
(\ref{anisotropic-Hamiltonian}), we will present below the
explicit analytical solutions for three cases. The first one is
$\boldsymbol{\mathfrak{N}} \parallel z$ and ${\bf B}
\parallel z$, in this case we obtain the following unitary evolution matrix
$\mathscr{U}_{{\rm Mu}^{\ast}zz } (t)$:
\begin{equation}
\label{unitary-evolution-anisotropic-zz}
\mathscr{U}_{{\rm Mu}^{\ast}zz} (t) = e^{i(a+d)t} \left(%
\begin{array}{cccc}
  e^{-i(2(a+d)-b_{-})t} & 0 & 0 & 0 \\
  0 & \cos ct + i (b_{+}/c) \sin ct & -i (2a/c) \sin ct & 0 \\
  0 & -i (2a/c) \sin ct & \cos ct - i (b_{+}/c) \sin ct & 0 \\
  0 & 0 & 0 & e^{-i(2(a+d)+b_{-})t} \\
\end{array}%
\right),
\end{equation}
\noindent where $d = \Delta A/(4\hbar)$.

In case $\boldsymbol{\mathfrak{N}} \parallel x$ and ${\bf B}
\parallel z$, we obtain symmetrical matrix $\mathscr{U}_{{\rm Mu}^{\ast}xz}
(t)$ with the elements
\begin{eqnarray}
&& \label{unitary-evolution-anisotropic-xz} [\mathscr{U}_{{\rm
Mu}^{\ast}xz} (t)]_{11} = e^{-iat} \big[
\cos ft+ i (b_{-}/f) \sin ft \big], \\
&& [\mathscr{U}_{{\rm Mu}^{\ast}xz} (t)]_{12} = [\mathscr{U}_{{\rm
Mu}^{\ast}xz} (t)]_{13} = [\mathscr{U}_{{\rm Mu}^{\ast}xz}
(t)]_{24} = [\mathscr{U}_{{\rm Mu}^{\ast}xz} (t)]_{34} = 0, \\
&& [\mathscr{U}_{{\rm Mu}^{\ast}xz} (t)]_{14} = -i e^{-iat}
(d/f) \sin ft , \\
&& [\mathscr{U}_{{\rm Mu}^{\ast}xz} (t)]_{22} = e^{iat} \big[ \cos
ht+ i (b_{+}/h) \sin ht \big], \\
&& [\mathscr{U}_{{\rm Mu}^{\ast}xz} (t)]_{23} = -i e^{iat}
[(2a+d)/h] \sin ht, \\
&& [\mathscr{U}_{{\rm Mu}^{\ast}xz} (t)]_{33} = e^{iat} \big[ \cos
ht - i (b_{+}/h) \sin ht \big], \\
&& [\mathscr{U}_{{\rm Mu}^{\ast}xz} (t)]_{44} = e^{-iat} \big[
\cos ft+ i (b_{-}/f) \sin ft \big],
\end{eqnarray}
\noindent where $f= \sqrt{(\Delta A)^2+4B^2(g_{\mu}\mu_{\mu} -
g_{e}\mu_{e})^2}/(4\hbar)$ and $h = \sqrt{(A+\Delta A
/2)^2+B^2(g_{\mu}\mu_{\mu} + g_{e}\mu_{e})^2}/(2\hbar)$.

Finally, in case $\boldsymbol{\mathfrak{N}} \parallel y$ and ${\bf
B} \parallel z$, we obtain the symmetrical unitary evolution
matrix $\mathscr{U}_{{\rm Mu}^{\ast}yz} (t)$ whose elements
coincide with elements of the matrix $\mathscr{U}_{{\rm
Mu}^{\ast}xz} (t)$ except for $[\mathscr{U}_{{\rm Mu}^{\ast}yz}
(t)]_{14} = - [\mathscr{U}_{{\rm Mu}^{\ast}xz} (t)]_{14}$.

\subsection{\label{subsection-Mu-like-Hamiltonian} Mu-like Hamiltonian}
\pst The muonium-like systems are described by Hamiltonians
(\ref{hyperfine-Hamiltonian})--(\ref{anisotropic-Hamiltonian}),
where the effective moment of the electron shell $j^{(e)}$ is
greater than $1/2$, i.e. the dimension of the operator $\hat{\bf
J}^{(e)}$ is greater than 2. In the particular case $j^{(e)}=1$,
the effective hyperfine interaction will result in the following
unitary evolution matrix:
\begin{equation}
\label{mu-like-hyperfine-evolution} \mathscr{U}_{\rm hf~\mu-like}
(t) = \left(
\begin{array}{cccccc}
 \mathscr{V}_1 & 0 & 0 & 0 & 0 & 0 \\
 0 & \mathscr{V}_{-} & 0 & \mathscr{V}_{2} & 0 & 0 \\
 0 & 0 & \mathscr{V}_{+} & 0 & \mathscr{V}_{2} & 0 \\
 0 & \mathscr{V}_{2} & 0 & \mathscr{V}_{+} & 0 & 0 \\
 0 & 0 & \mathscr{V}_{2} & 0 & \mathscr{V}_{-} & 0 \\
 0 & 0 & 0 & 0 & 0 & \mathscr{V}_1 \\
\end{array}
\right),
\end{equation}
\noindent where matrix elements $\mathscr{V}$ are expressed
through the strength of hyperfine interaction A
\begin{eqnarray}
&& \mathscr{V}_1 = e^{-iAt/(2\hbar)},\qquad \mathscr{V}_2 =
-ie^{iAt/(4\hbar)}(2\sqrt{2}/3)\sin(3At/(4\hbar)),\\
&& \mathscr{V}_{\pm} = e^{iAt/(4\hbar)}\big[ \cos(3At/(4\hbar))
\pm (i/3) \sin(3At/(4\hbar)) \big].
\end{eqnarray}
The entanglement of a Mu-like systems can be detected by using
tomogram $w(m^{(\mu)},{\bf n}^{(\mu)},m^{(e)},{\bf n}^{(e)})$
together with the qubit portrait method~\cite{chernega} and
Bell-like inequality (\ref{bell-like}). Moreover, PPT inequalities
(\ref{M-2})--(\ref{M-4}) are necessary conditions for
separability, so the non-zero value of quantity (\ref{measure}) is
a direct indicator of the entanglement of a Mu-like system as
well.

It is worth noting, that we restrict ourselves to the unitary
evolution of muonium-like systems and do not consider relaxation
processes. Their influence on system behavior and entanglement
will be discussed elsewhere.

\section{\label{section-tomography-Mu} Evolution and Entanglement of Mu-like Systems}
\pst In this section, we exploit the developed tomographic
approach to solve the problem of evolution and entanglement of Mu,
Mu$^{\ast}$, and Mu-like systems. We are going to indicate
entanglement of Mu-like systems and propose experimentally
accessible techniques for creation of entangled states. These two
points can be used in order to check experimentally Bell
inequalities with solid state spin states.

Let us consider the muonium atom.

\begin{figure}[t]
\begin{center}
\includegraphics{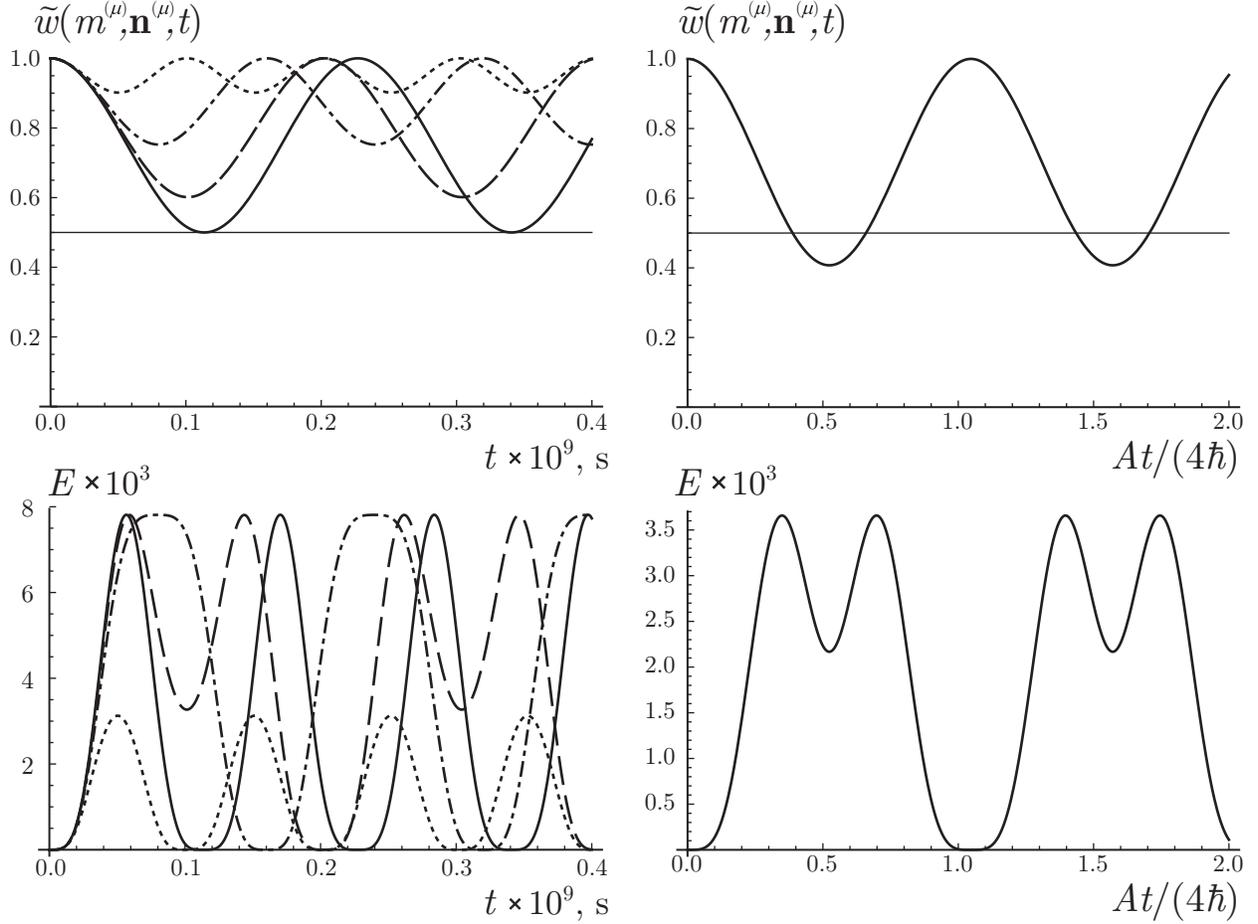}
\caption{\label{figure2} \small Time evolution of the reduced
tomogram $\widetilde{w}(1/2,{\bf n}^{(\mu)},t)$ of Mu in quartz
(left-up) and Mu-like system with effective electron moment
$j^{(e)}=1$ (right-up). Curved lines correspond to ${\bf
n}^{(\mu)} \parallel z$, horizontal (fine solid) lines correspond
to ${\bf n}^{(\mu)}
\parallel x$ and ${\bf n}^{(\mu)} \parallel y$. Time dependence of
entanglement for Mu in quartz (left-bottom) and for Mu-like system
(right-bottom). The initial state is chosen to be
$(2j^{(e)}+1)^{-1}|\Uparrow\rangle\langle \Uparrow | \otimes
\hat{I}^{(e)}$. Longitudinal magnetic field ${\bf B}
\parallel z$ with magnitude $B=0$ (heavy solid line), $B=790$~Gs
(dashed line), $B=1580$~Gs (chain line), and $B=3160$~Gs (dotted
line).}
\end{center}
\end{figure}

As it is shown in Sec.~\ref{section-evolution-tomogram}, for a
given Hamiltonian, the evolution is thoroughly determined by the
initial state of a system. In order to demonstrate how the
developed method works, we focus on the simplest and physically
clear initial state. Since the retarding time of muons in solids
is $\sim 10^{-10}$~s, the muons conserve their polarization and
are assumed $100\%$ polarized. In contrast, during the formation
of muonium the captured electrons have either up or down spin-$z$
projection, i.e. the electron subsystem is in the maximally mixed
state (non-coherent mixture). The reference time $t=0$ is the
formation of a muonium. The initial state of a muonium is
\begin{equation}
\label{initial-rho} \hat{\rho} (0) = |\Uparrow\rangle\langle
\Uparrow | \otimes \frac{1}{2} \hat{I}^{(e)}.
\end{equation}
\noindent Consequently, the initial individual two-spin tomogram
(\ref{individual-two-spin-tomogram}) reads
\begin{equation}
\label{initial-tomogram} w_0(m^{(\mu)},{\bf
n}^{(\mu)},m^{(e)},{\bf n}^{(e)}) = \frac{1}{2}\left( \frac{1}{2}
+ m^{(\mu)}{n}_z^{(\mu)} \right),
\end{equation}
\noindent that is the state is factorized (simply separable) and
tomogram (\ref{initial-tomogram}) does not depend on the variables
$m^{(e)}$ and ${\bf n}^{(e)}$ ascribed to the electron.
Substituting the initial tomogram (\ref{initial-tomogram}) for
$w_0(\widetilde{m}^{(\mu)},\widetilde{\bf
n}^{(\mu)},\widetilde{m}^{(e)},\widetilde{\bf n}^{(e)})$ in
(\ref{evolution-individual-tomogram}) and separating variables
$\{\widetilde{m}^{(\mu)},\widetilde{\bf n}^{(\mu)}\}$ and
$\{\widetilde{m}^{(e)},\widetilde{\bf n}^{(e)}\}$, we readily
obtain a solution in the form of the time-dependent tomogram
\begin{eqnarray}
\label{evolution-individual-tomogram-solution} w(m^{(\mu)},{\bf
n}^{(\mu)},m^{(e)},{\bf n}^{(e)},t) &=& \frac{1}{4} \Big[ 1 +
\langle j^{(\mu)}m^{(\mu)}, j^{(e)}m^{(e)} | \hat{R}({\bf
n}^{(\mu)}) \otimes \hat{R}({\bf n}^{(e)})
\nonumber\\
&& \times \hat{\mathscr{U}}(t) \hat{\sigma}_z^{(\mu)} \otimes
\hat{I}^{(e)} \hat{\mathscr{U}}^{\dag}(t) ~ \hat{R}^{\dag}({\bf
n}^{(\mu)}) \otimes \hat{R}^{\dag}({\bf n}^{(e)}) |
j^{(\mu)}m^{(\mu)}, j^{(e)}m^{(e)} \rangle \Big], \qquad\qquad
\end{eqnarray}
\noindent which is suitable to describe the evolution of muonium
governed by any unitary evolution (\ref{unitary-evolution-hf})--
(\ref{unitary-evolution-anisotropic-xz}).

Let us first consider the evolution in approximation of hyperfine
interaction with Hamiltonian (\ref{hyperfine-Hamiltonian}).

On substituting the unitary evolution matrix
(\ref{unitary-evolution-hf}) for $\mathscr{U}(t)$ in
(\ref{evolution-individual-tomogram-solution}), the direct
calculation yields
\begin{eqnarray}
\label{w-mu-e-hf-t} w(m^{(\mu)},{\bf n}^{(\mu)},m^{(e)},{\bf
n}^{(e)},t)
 &=& \frac{1}{4} \Big( 1+ m^{(\mu)}{n}_z^{(\mu)} +
m^{(e)}{n}_z^{(e)} + (m^{(\mu)}{n}_z^{(\mu)} - m^{(e)}{n}_z^{(e)})
\cos \omega_0 t \nonumber\\
&& + 2 m^{(\mu)} m^{(e)} [{\bf n}^{(\mu)} \times {\bf
n}^{(e)}]_{z} \sin \omega_0 t \Big).
\end{eqnarray}
As mentioned in Sec.~\ref{subsubsection-individual-2-spin-tom},
the MuSR-experiment enables us to measure the reduced tomogram
(tomogram of muon subsystem). In case of the hyperfine interaction
evolution, the reduced tomogram reads

\begin{equation}
\widetilde{w}(m^{(\mu)}, {\bf n}^{(\mu)},t) = \frac{1}{2} \left[
1+ m^{(\mu)}{n}_z^{(\mu)} (1+\cos \omega_0 t) \right].
\end{equation}

\begin{wrapfigure}{r}{82mm}
\begin{center}
\vspace{-8mm}
\includegraphics{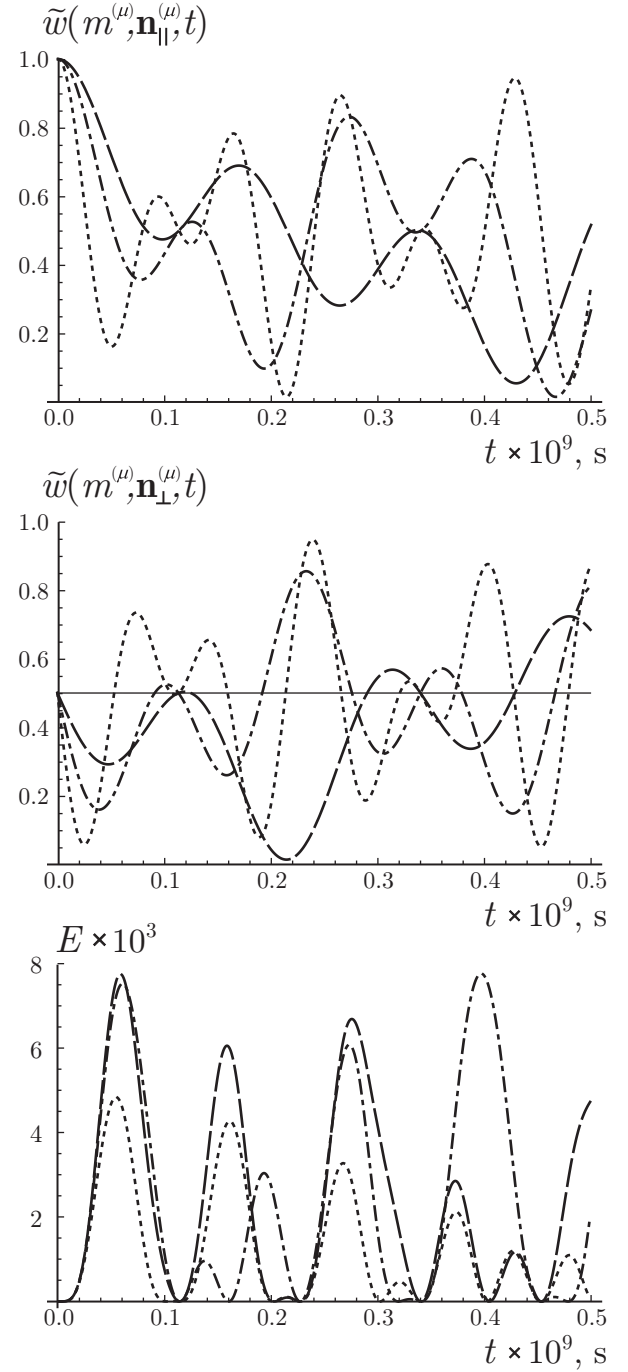}
\vspace{-3mm} \caption{\label{figure3} \small Time evolution of
the reduced tomogram $\widetilde{w}(1/2,{\bf n}^{(\mu)},t)$ of Mu
in quartz with ${\bf n}^{(\mu)} \parallel z$ (up), ${\bf
n}^{(\mu)}
\parallel x$ (middle, curved lines), and  ${\bf n}^{(\mu)}
\parallel y$ (middle, horizontal fine solid line). Time dependence of entanglement for Mu in
quartz (bottom). The initial state is $|\Uparrow\rangle\langle
\Uparrow | \otimes \frac{1}{2}\hat{I}^{(e)}$. Transversal magnetic
field ${\bf B}
\parallel x$ with magnitude $B=790$~Gs
(dashed line), $B=1580$~Gs (chain line), and $B=3160$~Gs (dotted
line).\vspace{-10mm}}
\end{center}
\end{wrapfigure}

The last term in formula (\ref{w-mu-e-hf-t}) is responsible for
entanglement of muon and electron spins. The direct calculation
shows that the maximum Bell-like number (\ref{bell-like}) for
tomogram (\ref{w-mu-e-hf-t}) is $\max_{{\bf n}_{1,2}^{(\mu,e)}} B
= | \sin \omega_0 t |$. Thus, the Bell number satisfies $|B| \le
1$ and is unappropriate for detecting entanglement of free
muonium. Partly, it is due to the initial state
(\ref{initial-rho})--(\ref{initial-tomogram}) which is obviously
mixed whereas the operation of the Bell-like number is best in
case of pure states. Applying the PPT criterion to the tomogram
(\ref{w-mu-e-hf-t}), we obtain $E=(1/128)\sin^4 \omega_0 t$, which
reveals the entanglement of a free muonium at the time moments $t
\ne \pi k / \omega_0$, $k=0,1,2,\ldots$ Time evolution of the
reduced tomogram and the entanglement measure is visualized in
Fig. \ref{figure2} (muonium in quartz, zero field).

Dealing with more complicated systems, for the sake of brevity, we
will omit analytical solutions of the tomographic evolution
equation (\ref{evolution-individual-tomogram}) and calculations of
the entanglement measure (\ref{measure}).

In Fig. \ref{figure2}, we present the reduced (muon) tomogram and
the entanglement measure for the simplest Mu-like system ``muon +
spin 1'' which is governed by unitary evolution
(\ref{mu-like-hyperfine-evolution}).

As far as muonium in quartz is concerned, we exploit the
Hamiltonian (\ref{isotropic-Hamiltonian}) with $A=(2\pi\hbar)
\cdot 4404$~MHz and different magnetic fields
$B=0,B_c/2,B_c,2B_c$, where $B_c=1580$~Gs is a critical field,
i.e. $B(g_{e}\mu_{e} - g_{\mu}\mu_{\mu}) = A$. In Fig.
\ref{figure2}, we illustrate the behavior of the observed reduced
tomogram $\widetilde{w}(m^{(\mu)}, {\bf n}^{(\mu)},t)$ and the
entanglement measure $E$ when ${\bf B} \parallel z$ (longitudinal
field). While increasing the magnetic field intensity $B$, the
tomographic value $\widetilde{w}(m^{(\mu)}=1/2, {\bf
n}_z^{(\mu)},t)$ oscillates with increasing frequency and tends to
1 due to a preferable muon spin polarization along positive
$z$-axis and electron spin polarization along negative $z$-axis
(separable state). For the same reason, the entanglement measure
$E$ decreases substantially when the strength of magnetic filed is
greater than the critical value $B_c$. The probability to obtain
muon spin projection $+1/2$ along $x$- or $y$-axis is exactly
$1/2$ and is not sensitive to the magnitude of magnetic field
aligned along $z$-axis.

The case of muonium in quartz, ${\bf B}
\parallel x$ (transversal field) is depicted in Fig. \ref{figure3}. Both tomographic values $\widetilde{w}(1/2, {\bf
n}_z^{(\mu)},t)$ and $\widetilde{w}(1/2, {\bf n}_x^{(\mu)},t)$
exhibit a rather complicated dynamics, whereas $\widetilde{w}(1/2,
{\bf n}_y^{(\mu)},t) = 1/2$. Again, the essential decrease of
entanglement is observed when $B > B_c$.

\begin{figure}
\begin{center}
\includegraphics{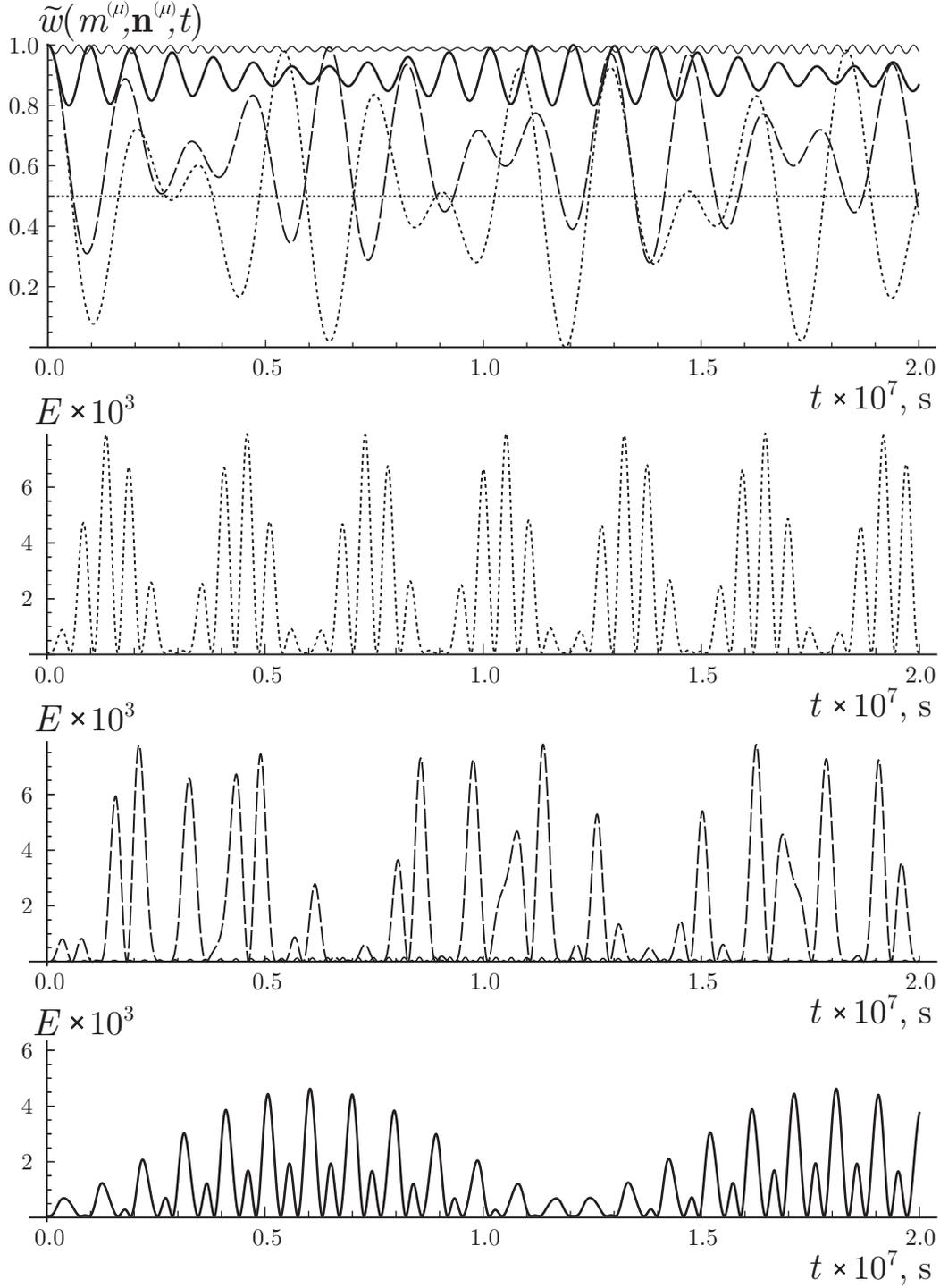}
\caption{\label{figure4} \small Time evolution of the reduced
tomogram $\widetilde{w}(1/2,{\bf n}^{(\mu)},t)$ of Mu$^{\ast}$ in
Si with ${\bf n}^{(\mu)} \parallel z$ (up, curved lines), ${\bf
n}^{(\mu)}
\parallel x$  and  ${\bf n}^{(\mu)}
\parallel y$ (up, horizontal fine dotted line). Time dependence of entanglement for Mu$^{\ast}$ in
Si (three bottom figures). The initial state is
$|\Uparrow\rangle\langle \Uparrow | \otimes
\frac{1}{2}\hat{I}^{(e)}$. Anisotropy direction
$\boldsymbol{\mathfrak{N}}
\parallel x$. Longitudinal magnetic field ${\bf B}
\parallel z$ with magnitude $B=0$ (dotted line), $B=10$~Gs
(dashed line), $B=33$~Gs (heavy solid line), and $B=100$~Gs (fine
solid line).}
\end{center}
\end{figure}

Anomalous muonium behaves as an ordinary muonium if
$\boldsymbol{\mathfrak{N}} \parallel z$ and ${\bf B}
\parallel z$, with the only difference being in time scale. The case $\boldsymbol{\mathfrak{N}} \parallel x$ and ${\bf B}
\parallel z$ is illustrated in Fig. \ref{figure4}, we depict the reduced tomogram and entanglement evolution
for anomalous muonium in silicon ($A=92.595$~MHz, $\Delta A =
-75.776$~MHz~\cite{patterson79,patterson88}) in presence of
different magnetic fields $B=0$, $10$, $33$, and $100$~Gs
(critical field $B_c=33$~Gs). A two-frequency modulation of the
reduced tomogram and a decrease of entanglement are observed if $B
\ge B_c$.

\subsection{\label{subsection-reconstruction} Reconstruction of the Initial State}
\pst The strong dependence of the reduced (muon) spin tomogram on
time can serve as an indicator of a system Hamiltonian and allows
to determine numerical values of parameters ($A$, $\Delta A$,
${\bf B}$, $\boldsymbol{\mathfrak{N}}$). Given the evolution
operator $\hat{\mathscr{U}}(t)$ and the measured reduced tomogram
$\widetilde{w}(m^{(\mu)}, {\bf n}^{(\mu)},t)$, it is possible, in
general, to reconstruct the initial state
$\hat{\rho}(0)$~\cite{measuring-dens-matrix}. Indeed, fifteen
independent values $\{\widetilde{w}(1/2, {\bf
n}_k^{(\mu)},t_l)\}$, $k=x,y,z$, $l=1,2,\ldots,5$ allow the
reconstruction of fifteen independent real parameters of $4\times
4$ density matrix $\rho(0)$. The requirement on time moments
$\{t_l\}_{l=1}^{5}$ is to avoid a periodical coincidence. The
important requirement on unitary evolution $\hat{\mathscr{U}}(t)$
is to break an internal symmetry of the two-spin state, i.e. an
anisotropic term like $\Delta A$ in
(\ref{anisotropic-Hamiltonian}) is to be presented in the system
Hamiltonian. It is worth mentioning that the idea of using time
unitary evolution to measure a symplectic tomography of
Bose-Einstein condensate was successfully applied in the
paper~\cite{campo-manko}.

\section{\label{section-conclusions} Conclusions}
\pst To conclude we point out the main results of our work.

We applied the tomographic probability approach to the problem of
muon spin rotation/relaxation/ resonance experiments. The relation
between MuSR histogram and muon spin tomogram was established. In
view of this, all MuSR experiments can be interpreted as measuring
the reduced tomogram of a quantum state of two-spin system,
namely, muon-electron one. The density matrix of this system was
mapped onto a joint probability distribution of muon and electron
spin projections onto their quantization directions. Such a
tomogram was shown to completely describe any muon-electron
quantum state. The reduced density matrix describing the muon
state was mapped bijectively onto a spin-tomographic probability
distribution of the muon spin projection onto a chosen direction
(muon spin tomogram).

Time evolution of muon-electron system was investigated within the
framework of the tomographic representation. The entanglement
phenomenon between muon and electron spins was shown to appear due
to different interactions including the hyperfine interaction, the
anisotropic hyperfine interaction and the interaction of muon and
electron magnetic moments with an external magnetic field. We
studied the behavior of the Bell-like tomographic inequality and
showed that, for a discussed initial system state, there is no
violation of this inequality. The PPT criterion of separability
was discussed in the tomographic probability representation. The
negativity indicator of entanglement was applied to the
muon-electron system. Using such a tomographic entanglement
measure, we studied the influence of different kinds of the
discussed interactions onto the entanglement phenomenon. The
detailed analysis of entanglement under the influence of
dissipation within the framework of kinetic equation with
relaxation terms is a problem for further consideration.

\section*{Acknowledgments}
\pst The authors thank the Ministry of Education and Science of
the Russian Federation and the Federal Education Agency for
support under Project No. 2.1.1/5909 and Contract No. $\Pi$558
``Application of MuSR method to the investigation of
nanostructures." S.N.F. and V.I.M. are grateful to the Russian
Foundation for Basic Research for partial support under Projects
Nos. 09-02-00142 and 10-02-00312. The authors thank Igor
Chernousov for interesting comments.

\end{document}